\documentclass[letterpaper]{IEEEtran}
\IEEEoverridecommandlockouts
\usepackage{etex}

\IEEEoverridecommandlockouts                              
\pdfoutput=1

\usepackage{times}
\usepackage{latexsym,amsfonts,amsbsy,amssymb}
\usepackage{amsmath}
\usepackage{graphicx}
\usepackage{cite}
\usepackage{algorithm}
\usepackage{algorithmic}
\usepackage{multirow}
\usepackage{enumerate}
\usepackage{gensymb}
\usepackage{mathrsfs}
\usepackage{stmaryrd}
\usepackage{tabularx,booktabs}
\usepackage{comment}
\usepackage[table]{xcolor}
\usepackage[curve]{xypic}
\usepackage{subfig}
\usepackage{caption}
\usepackage{color}
\usepackage{tikz}
\usetikzlibrary{automata,positioning,shapes,arrows}

\newtheorem{definition}{Definition}

\newtheorem{remark}{Remark}
\newtheorem{example}{Example}
\newtheorem{problem}{Problem}

\usepackage{float}

\usepackage[algo2e]{algorithm2e}

\newcommand{\hashed}[1]{\ignorespaces}

\begin{document}
\title{Distributed Communication-aware Motion Planning for Networked Mobile Robots under Formal Specifications 
}
\author{Zhiyu~Liu,~\IEEEmembership{Student~Member,~IEEE,}
	    Bo~Wu,~\IEEEmembership{Member,~IEEE,}
        Jin~Dai,~\IEEEmembership{Student~Member,~IEEE,}
        and Hai~Lin~\IEEEmembership{Senior~Member,~IEEE}
        \thanks{The partial support of the National Science Foundation (Grant No. CNS-1446288, ECCS-1253488, IIS-1724070) and of the Army Research Laboratory (Grant No. W911NF- 17-1-0072) is gratefully acknowledged.}
\thanks{Z. Liu, J. Dai,  and H. Lin are with the Department of Electrical Engineering, University of Notre Dame, Notre Dame, IN 46556, USA, {\tt\small zliu9@nd.edu; jdai1@nd.edu; hlin1@nd.edu}.}
\thanks{B. Wu was with the University of Notre Dame, he is now with the Institute for Computational Engineering and Sciences (ICES), the University of Texas at Austin, Austin, TX 78712, USA, {\tt\small bwu3@utexas.edu}. }}
\date{}

\maketitle
\thispagestyle{empty}
\pagestyle{empty}

\begin{abstract}
Control and communication are often tightly coupled in motion planning of networked mobile robots, due to the fact that robotic motions will affect the overall communication quality, and the quality of service (QoS) of the communication among the robots will in turn affect their coordination performance. In this paper, we propose a control theoretical motion planning framework for a team of networked mobile robots in order to accomplish high-level spatial and temporal motion objectives while optimizing communication QoS. Desired motion specifications are formulated as Signal Temporal Logic (STL), whereas the communication performances to be optimized are captured by recently proposed Spatial Temporal Reach and Escape Logic (STREL) formulas. Both the STL and STREL specifications are encoded as mixed integer linear constraints posed on the system and/or environment state variables of the mobile robot network, where satisfactory control strategies can be computed by exploiting a distributed model predictive control (MPC) approach. To the best of the authors' knowledge, we are the first to study controller synthesis for STREL specifications. A two-layer hierarchical MPC procedure is proposed to efficiently solve the problem, whose soundness and completeness are formally ensured. The effectiveness of the proposed framework is validated by simulation examples.

\end{abstract}

\begin{IEEEkeywords}
Networked mobile robots, formal methods, motion planning, spatial temporal logic, optimization, model predictive control.
\end{IEEEkeywords}

\section{Introduction}

\IEEEPARstart{M}{otion} planning and control of mobile robots have drawn a considerable amount of research interest in recent years for various purposes, ranging from persistent surveillance \cite{smith2011optimal}, formation control \cite{nguyen2017formation}, target tracking \cite{foderaro2018distributed}  to simultaneous localization and mapping (SLAM) \cite{durrant2006simultaneous}. The classical motion planning problem aims at steering a mobile robot from an initial configuration to some final configurations while avoiding collision with any obstacles along the way. Many computationally efficient planning methods have been proposed in this context to compute such collision-free paths, see, e.g., \cite{choset2005principles,lavalle2006planning,belta2007symbolic,fainekos2009temporal,karaman2011sampling} and the references therein. In addition to motion planning for a single robot, many contributions have also been made to solving motion planning problems for multiple and networked mobile robots under either a global specification \cite{chen2012formal,chen2013temporal,ulusoy2013optimality,lindemann2017decentralized} or a series of individual motion specifications \cite{guo2015multi,guo2016communication,tumova2016multi}. 

Inter-robot communication often plays an important role in motion planning of multi-robot systems as it provides robots with essential information for decision-making \cite{khan2015information}. The communication quality of service (QoS), which is often influenced by package dropout, path loss, fading and/or shadowing effects during the signal propagation, is tightly coupled with the motion performance that determines the spatial deployment of the robots \cite{yan2014go}. Communication QoS must be taken into consideration for tasks such as emergency response, persistent surveillance, search and rescue, and fusion of mobile sensor networks, where high quality sensing data, such as visual data or point cloud, are required for real-time decision-making \cite{muralidharan2017energy}. 

The importance of communication hence inspires the pursuit of communication-aware motion planning of mobile robots. Control polices based on algebraic graph theory were applied to maintain the connectivity of the robotic network \cite{ji2007distributed,zavlanos2011graph,poonawala2015collision}, which can be characterized by the second-smallest eigenvalue (Fiedler eigenvalue) of the communication graph's Laplacian matrix. 
However, network connectivity is a global property that does not characterize the QoS between any two communicating robots. Grancharova et al. \cite{grancharova2015uavs} considered co-optimization of motion and communication by proposing a distributed model predictive control (MPC) method to compute a satisfactory trajectory for each robot such that constraints on communication channel capacity were respected. Nevertheless, channel capacity does not directly reflect the current QoS as it is a theoretical upper bound of the achievable data rate. An optimal relay node placement problem was studied in \cite{wu2014channel}, where mobile robots inspecting a pipeline could reliably communicate with base stations; however, the relay nodes therein were assumed to be static, which limited the flexibility of the robot network to various environments. Mobile routers were considered in \cite{el2013mobile}, where energy consumption by motion and communication were minimized without considering the QoS. Furthermore, the sensing nodes were assumed to be static; whereas in many practical applications, the robots are required to cover a wide range of areas for exploration. Mobile relays which minimize the bit error rate between two communication nodes were studied in \cite{yan2012robotic}, where sensing nodes were also assumed to be static. Wu et al. \cite{wu2017energy} studied the communication-aware motion planning in outdoor scenarios by accounting for the static base station, mobile relay, and sensing robots, where the trajectories of sensing robots were restrained. Yan and Mostofi \cite{yan2014go} studied a motion-communication co-optimization problem of a mobile robot. However, the trajectory of the robot was pre-defined and the communication performance was optimized by controlling the robot's motion velocity, transmission rate and stop time, where a more recent paper studies a similar problem with multi-robot systems and online adaptation \cite{ali2018motion}. More complex channel models, such as the Gaussian process model, was considered in \cite{fink2013robust}. Nevertheless, obtaining a reliable model can be time and data consuming for complex environments, and is not suitable for large-scale multi-robot systems. 

Motivated by the aforementioned concerns, we are interested in designing motion controllers for each robot of a given team of networked mobile robots in a shared environment with communication base stations such that: (i) each robot can satisfy its own motion specifications while avoiding collisions with other robots or unsafe regions; (ii) the communication QoS between the sensing robots and the base stations can be optimized. To this end, we use Signal Temporal Logic (STL) \cite{maler2004monitoring,liuacc} formulas to describe local motion and safety requirements for each robot, and Spatial Temporal Reach and Escape Logic (STREL) \cite{bartocci2017monitoring} formulas to represent communication QoS requirements. A mixed-integer linear programming (MILP) formalism is established to encode both STL and STREL formulas as constraints on the state and environment variables of a joint motion-communication co-optimization problem. Specifically, we consider an indoor scenario with different communication channel characteristics and missions that require communication with a high data rate. The desired trajectories for the robots are computed online by employing a distributed MPC architecture. Our work differs from the prior work \cite{wu2014channel,el2013mobile,yan2012robotic,wu2017energy} in the sense that we allow relay and sensing robots to be both mobile and their trajectories are neither pre-defined nor restrained. Furthermore, we assume all the robots and the base stations are equipped with a millimeter wave communication system for high data throughput. Since the signal in millimeter wave communication is highly directional and has high reflection losses \cite{torkildson2009millimeter}, our proposed channel model considers the path loss as a function of distance and whether line of sight communication exists. This distinguishes our work from the disc model in the literature, which assumes perfect communication within a certain distance \cite{zavlanos2009hybrid,tekdas2010maintaining,wu2015combined}. 

Compared to our previous conference publication \cite{liu2017distributed}, we use the recently proposed STREL rather than the SpaTeL \cite{haghighi2015spatel} formulas to better specify local communication requirements, due to the fact that STREL formulas that specify the property at each location and time shall depend only on the neighbors of a given robot, while SpaTeL formulas are less flexible. To reduce the computational complexity for communication-aware motion planning, we employ a distributed MPC framework in \cite{liu2017distributed}. Although completeness of such an MPC technique cannot be guaranteed in some cases \cite{watterson2015safe}, we propose a two-layer hierarchical structure of distributed MPC which permits appropriate motion controllers to be synthesized locally while completeness of this approache can be guaranteed. The main contributions of this paper are summarized as follows.

(i) Given a team of networked mobile robots moving in a shared environment with communication base stations, we solve the communication and motion co-optimization problem for the robots by constructing a local controller for each robot such that it steers the robot to fulfill local motion specifications specified as STL formulas and to obey local communication constraints captured by STREL formulas.

(ii) We propose a Boolean encoding scheme for STREL formulas so that the controller synthesis problem can be translated into an MILP, which can be solved efficiently. To the best of the authors' knowledge, this paper is the first one to tackle controller synthesis under STREL specifications. 

(iii) We develop a distributed two-layer hierarchical control framework to achieve a scalable co-optimization algorithm for networked mobile robots. The computation complexity is reduced significantly compared to a monolithic approach. The high layer of the proposed framework provides a set of waypoints to guide the low layer so that the completeness can be guaranteed. The low layer implements controller synthesis for the given STL and STREL specifications under the MPC framework, and the resulting trace is guaranteed to satisfy both the motion and communication requirements.  

The remainder of this paper is organized as follows. In Section II, we briefly introduce the necessary preliminaries of STL and STREL as well as the models of multiple mobile robots under consideration. The communication-aware motion planning problem for the multi-robot network under STL-STREL specifications is then formally formulated in Section III. In Section IV, we propose the MILP encoding schemes for STL and STREL specifications. Based on the obtained MILP constraints, a distributed MPC control framework is exploited with guarantee of completeness to synthesize local motion controllers in Section V. The correctness of our proposed planning framework is validated in Section VI through simulation examples. We conclude this paper in Section VII.

{\it Notations} The notations used throughout this paper are fairly standard. $\mathbb{R}$, $\mathbb{N}$ and $\mathbb{B}$ denote the set of real numbers, the set of natural numbers and the set of binary bits $\{0,1\}$, respectively. $\mathbb{R}^n$ denotes the $n$-dimensional Euclidean space and $\mathbb{R}^{n\times m}$ denotes the set of $n\times m$ real matrices. For a given vector or matrix $A$, $A^T$ denotes its transpose and its 2-norm is denoted as $\|A\|$. For a given set $\mathcal{S}$, we let $2^\mathcal{S}$ and $|\mathcal{S}|$ denote the power set and the cardinality of $\mathcal{S}$, respectively. In addition, $\mathcal{S}^\omega$ denotes the set of infinite-length strings whose elements are drawn from $\mathcal{S}$. Finally, for $m,n\in\mathbb{N}$ with $m\le n$, we use the notation $[m,n]$ to denote the set of consecutive integers $\{m,m+1,\ldots,n\}$.

\section{Preliminaries and Models}

\subsection{Robot Dynamics}

We assume that the team of networked mobile robots under consideration in this paper is composed of $P$ robots performing in a shared 2-D environment with heterogeneous dynamics and unique identities, namely $\mathcal{P}=\{1,2, \ldots, P\}$. Some robots are equipped with various mission execution capabilities to explore certain areas of interest while others are deployed as communication relay robots. For each $i\in \mathcal{P}$, the evolution of the robot $\mathcal{R}_i$ is governed by the following linear dynamics
\begin{equation}
\dot x_i(t)=A_ix_i(t)+B_iu_i(t),
\label{dynamic}
\end{equation}
Here, $x_i=[p_i^T\quad v_i^T \quad a_i^T]^T$ is the state of robot $\mathcal{R}_i$, where $p_i, v_i\in\mathbb{R}^2$ are the position and velocity of the robot, respectively, and $a_i\in\mathbb{B}^2$ is a vector that encodes different attributes of the robot. Since we consider three different type of robots defined in the following sections, a 2-dimensional binary vector is sufficient to determine a specific attribute. Additionally, $u_i=[u_{i,1}\quad u_{i,2}]^T\in \mathcal{U}\subseteq \mathbb{R}^2$ is the local control inputs, where $\mathcal{U}$ stands for the set of admissible controls, and $x_i(0)=x_{i,0}$ is the initial state. $(A,B)$ is a controllable pair of matrices with proper dimensions. The environment $\mathcal{X}$ is given by a convex polygonal subset of the $2$-D Euclidean space $\mathbb{R}^2$. Let $\mathcal{X}_{obs}\subseteq \mathcal{X}$ be the regions in the environment occupied by polygon obstacles. $\mathcal{X}_{free}=\mathcal{X}\setminus\mathcal{X}_{obs}$ denotes the obstacle-free working space for the multi-robot system.

To run the distributed communication-aware motion planning in an online manner inspired by \cite{raman2015reactive}, we assume that, given an appropriate sampling time $\Delta t>0$, the continuous-time model (\ref{dynamic}) of the robot $\mathcal{R}_i$ admits a discrete-time approximation of the following form:

\begin{equation}
x_i(t_{k+1})=A_{i,d}x_i(t_{k})+B_{i,d}u_i(t_{k}),
\label{agentdynamic}
\end{equation}
where $k\in\mathbb{N}$ is the sampling index and $\Delta t$ is selected such that $(A_{i,d}, B_{i,d})$ is also controllable. The sampling is uniformly performed, i.e., for each $k>0$, $t_{k+1}-t_k=\Delta t$.

Given $x_{i,k}$ and $\bf{u}_i\in \mathcal{U}^\omega$, $i\in\mathcal{P}$, a (state) {\it run} ${\bf x}_i= x_{i,k}x_{i,k+1}x_{i,k+2}\ldots\in(\mathbb{R}^4\times\mathbb{B}^2)^\omega$ generated by the robot $\mathcal{R}_i$ (\ref{agentdynamic}) with a sequence of control inputs $\bf {u_i}$ is an infinite sequence obtained from $\mathcal{R}_i$'s state trajectory, where $x_{i,k}=x_i(t_k)$ is the state of the system at time index $k$, and for each $k\in\mathbb{N}$, there exists a control input $u_{i,k}=u_i(t_k)\in \mathcal{U}$ such that $x_i(t_{k+1})=A_{i,d}x_i(t_{k})+B_{i,d}u_i(t_{k})$. Under the MPC framework with planning horizon $H\in\mathbb{N}$ (cf. Section III), given a local state $x_{i,k}$ and a finite sequence of local control inputs ${\bf u}^H_i = u_{i,k}u_{i,k+1}u_{i,k+2}\ldots u_{i,k+H-1}$, the resulting horizon-$H$ run of the robot $\mathcal{R}_i$, written as ${\bf x}_i(x_{i,k},{\bf u}^H_i)=x_{i,k}x_{i,k+1}x_{i,k+2}\ldots x_{i,k+H-1}$, is unique.

\subsection{Inter-robot Communication Models}
Typically, networked robots may be assigned to different roles and responsibilities, and inter-robot communication is required to ensure proper coordination between them for safety and efficient mission execution. Furthermore, reliable communication is also needed between the robots and base stations to collect the  environment data and let base stations provide global information services such as clock synchronization for robots, which is essential for distributed algorithms. Therefore, we explicitly consider communication as an optimization objective.

\begin{figure}[h]
	\centering
	\includegraphics[width=0.95\linewidth]{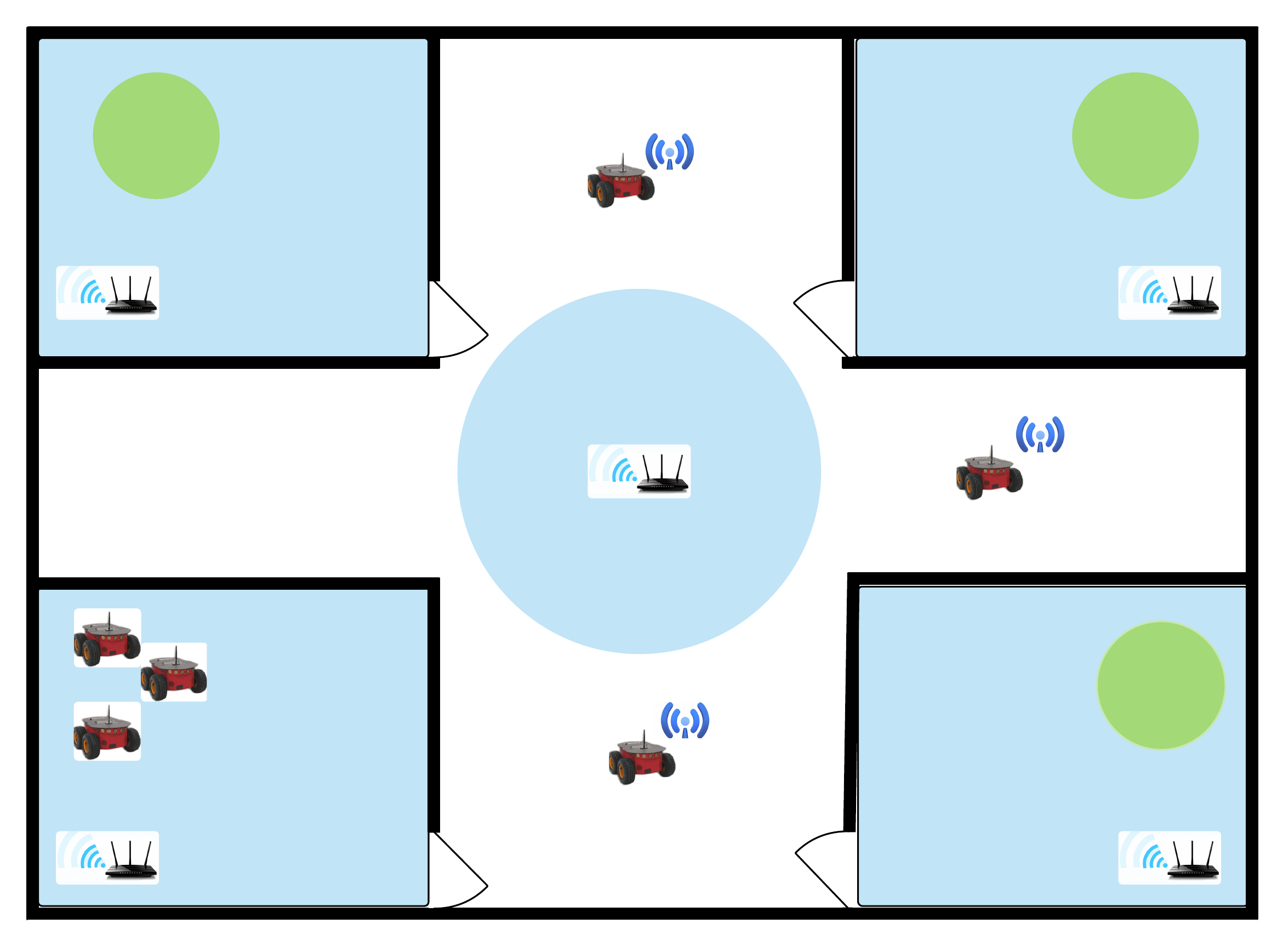}
	\caption{The indoor environment shared by the robots.}
	\label{comm}
	\vspace{-7.5mm}
\end{figure}

The QoS of inter-robot communication and the communication between robots and base stations are assumed to be subject to path loss and shadowing effect \cite{molisch2012wireless} due to obstacles like walls. In this paper, we consider a team of robots deployed in an indoor environment shown in Fig. \ref{comm}. Four rooms are located at the corners of the working space, each with one door opened to the hall. Each room is equipped with a static communication base station which is capable of covering the whole room. One extra base station is deployed at the center of the hall. Due to path loss and shadowing effect in the wireless millimeter wave communication channel \cite{niu2015survey}, only robots in the blue area can reach its corresponding base station. We assume that the base stations can communicate with each other since they are static and connected via cable. Initially, a team of robots is deployed in the room at the lower left corner. Their mission is to reach and collect information in the green area of other rooms. Several mobile robots with communication relay capability are placed in the hall to help other robots reach base stations in case they are not in the blue area.  

\subsection{Signal Temporal Logic}
We consider motion planning objectives that are specified by STL formulas, which are defined as follows.

\begin{definition}[STL Syntax]\rm
STL formulas are defined recursively as:
$$
\varphi::={\rm True}|\pi^\mu|\neg\pi^{\mu}|\varphi\land\psi|\varphi\lor\psi|\Box_{[a,b]} \psi | \varphi\sqcup_{[a,b]} \psi,
$$
\end{definition}
where $\pi^\mu$ is an atomic predicate $\mathbb{R}^n\to\{0,1\}$ whose truth value is determined by the sign of a function $\mu:\mathbb{R}^n\to\mathbb{R}$, i.e., $\pi^\mu$ is true if and only if $\mu({\bf x})>0$; and $\psi$ is an STL formula. The ``eventually" operator $\Diamond$ can also be defined here by setting $\Diamond_{[a,b]} \varphi={\rm True}\sqcup_{[a,b]} \varphi$.

The semantics of STL with respect to a discrete-time signal $\bf x$ are introduced as follows, where $({\bf x},t_k)\models \varphi$ denotes for which signal values and at what time index the formula $\varphi$ holds true.
\begin{definition}[STL Semantics]\rm
The validity of an STL formula $\varphi$ with respect to an infinite run ${\bf x}= x_0x_1x_2\ldots$ at time $t_k$ is defined inductively as follows.
\begin{enumerate}
\item $({\bf x},t_k)\models \mu$, if and only if $\mu(x_k)>0$;
\item $({\bf x},t_k)\models \neg\mu$, if and only if $\neg(({\bf x},t_k)\models \mu)$;
\item $({\bf x},t_k)\models \varphi\land\psi$, if and only if $({\bf x},t_k)\models \varphi$ and $({\bf x},t_k)\models \psi$;
\item $({\bf x},t_k)\models \varphi\lor\psi$, if and only if $({\bf x},t_k)\models \varphi$ or $({\bf x},t_k)\models \psi$;
\item $({\bf x},t_k)\models \Box_{[a,b]}\varphi$, if and only if $\forall t_{k'}\in[t_k+a,t_k+b]$, $({\bf x},t_{k'})\models \varphi$;
\item $({\bf x},t_k)\models \varphi\sqcup_{[a,b]}\psi$, if and only if $\exists t_{k'}\in[t_k+a,t_k+b]$ such that $({\bf x},t_{k'})\models \psi$ and $\forall t_{k''}\in[t_k,t_{k'}]$, $({\bf x},t_{k''})\models \varphi$;
\item $({\bf x},t_k)\models \Diamond_{[a,b]}\varphi$, if and only if $\exists t_{k'}\in[t_k+a,t_k+b]$, $({\bf x},t_{k'})\models \varphi$.
\end{enumerate}
\end{definition}

A run ${\bf x}$ satisfies $\varphi$, denoted by $\bf x\models\varphi$, if $({\bf x},t_0)\models\varphi$.

Intuitively, $\bf x\models \Box_{[a,b]}\varphi$ if $\varphi$ holds at every time step between $a$ and $b$, ${\bf x}\models \varphi\sqcup_{[a,b]}\psi$ if $\varphi$ holds at every time step before $\psi$ holds and $\psi$ holds at some time step between $a$ and $b$, and ${\bf x}\models \Diamond_{[a,b]}\varphi$ if $\varphi$ holds at some time step between $a$ and $b$.

An STL formula $\varphi$ is {\it bounded-time} if it contains no unbounded operators. The bound of $\varphi$ can be interpreted as the horizon of future predicted signals $\bf x$ that is needed to calculate the satisfaction of $\varphi$.

In addition to STL syntax and semantics, \cite{raman2014model} introduced space robustness (robustness for short) of satisfaction of an STL formula in a quantitative manner. Specifically, the robustness assigned a real-valued measure $\rho^\varphi$ of a signal $\bf x$ at $t$ such that $({\bf x},t)\models \varphi$ if and only if $\rho^\varphi({\bf x},t)>0$.

\begin{definition}[Space Robustness]\rm
\cite{raman2014model} The space robustness of an STL formula's satisfaction is defined as:
\begin{enumerate}
\item $\rho^\mu({\bf x},t)=\mu({\bf x},t)$;
\item $\rho^{\neg\mu}({\bf x},t)=-\mu({\bf x},t)$;
\item $\rho^{\varphi\land\psi}({\bf x},t)=\min(\rho^\varphi({\bf x},t), \rho^\psi({\bf x},t))$;
\item $\rho^{\varphi\lor\psi}({\bf x},t)=\max(\rho^\varphi({\bf x},t), \rho^\psi({\bf x},t))$;
\item $\rho^{\Box_{[a,b]}\varphi}({\bf x},t)=\min_{t'\in[t+a,t+b]} \rho^\varphi({\bf x},t')$;
\item $\rho^{\varphi\sqcup_{[a,b]}\psi}({\bf x},t)= \max_{t'\in[t+a,t+b]}(\min(\rho^\psi({\bf x},t), \\ \min_{t''\in[t,t']} \rho^\varphi({\bf x},t'')))$,
\end{enumerate}
\end{definition}
where $\rho^\mu$ is an abbreviation for $\rho^{\pi^\mu}$.

The robustness of satisfaction for an STL formula is computed recursively from the above semantics in a straightforward manner by propagating the values of the functions associated with each operand using $\min$ and $\max$ operators corresponding to various STL operators. For example, the robust satisfaction of $\mu_1\equiv x-3>0$ at time $t=0$ is $\rho^{\mu_1}=x_0-3$. Temporal operators are treated as conjunctions and disjunctions along the time axis.

With the employment of the $\max$ and $\min$ operations, space robustness $\rho^\varphi({\bf x},t)$ characterizes ``how much" a run $\bf x$ satisfies $\varphi$ by considering the weakest points along $\bf x$ at which $\varphi$ is least satisfied.

\subsection{Spatial Temporal Reach and Escape Logic}

A novel spatial temporal logic, Spatial Temporal Reach and Escape Logic (STREL), has been recently proposed in \cite{bartocci2017monitoring} to describe emergent behaviors of robotic systems. It enables the specification and monitoring of spatial temporal requirements during the execution of mobile and spatially distributed multi-robot systems such as mobile ad-hoc sensor networks and swarming robotics. STREL is defined by extending STL with two novel spatial operators, ``reach" and ``escape". Unlike other spatial temporal logic such as SpaTeL \cite{haghighi2016robotic}, the satisfaction of the property at each location and time step depends locally only on the satisfaction of its neighbours. Similar to STL, its semantics can be either qualitative, ranging over Boolean values, or quantitative, ranging over real values, which gives its ability to characterize the robustness of the multi-robot systems.

STREL uses a graph $G=(V,W)$ as the spatial model, where locations of the robot $\mathcal{R}_i$ $(i\in\mathcal{P})$ are treated as nodes $v_i\in V$ and the connection to another robot forms a weighted edge of the graph $w_{i,j}\in W$ for $\forall v_i,v_j\in V$. Depending on different applications, the weight maps the connection between two nodes into a different domain. For example, in a communication network that is captured by disc model where the connection among nodes is either on or off, the weight $w_{i,j}$ maps the relationship between two nodes into a binary domain $W: V\times V\rightarrow \mathbb{B}$. For some swarming robotics applications where the connection among robots is subject to the Euclidean distance, the weight maps the relation into a real domain $W: V\times V \rightarrow \mathbb{R}$. The flexibility in the definition of a spatial model using graph gives STREL an ability to specify a wide range of applications for networked robots.

Similar to the run in STL, a spatial temporal trace in STREL is defined based on the spatial model by using the graph \cite{bartocci2017monitoring}.
\begin{definition}[Spatial temporal Trace]
Let V be the set containing all nodes. A spatial temporal trace is a function for all $v_i\in V$ and $t\in \mathbb{T}=[0,T]$
\begin{align*}
    \overrightarrow{x}: V\rightarrow\mathbb{T}\rightarrow D^n
\end{align*}
where $D$ represents the domain of the trace such as Boolean domain and real domain. 
\end{definition}

Due to mobility in the multi-robot systems, spatial topology will change over time. Therefore, a location service function, defined below, is needed to return the spatial topology over time.

\begin{definition}[Location service]
A location service is a function $\lambda: V\times \mathbb{T}\rightarrow G$ which returns a spatial model $G$ for each $t$ in $[0,T]$.
\end{definition}

The following example illustrates the intuitive idea of the definition of spatial model $G$, trace $\overrightarrow{x}$ and location service function $\lambda$.

\begin{example}
Consider a heterogeneous multi-robot system shown in Fig. \ref{exp} with three types of robots. Base stations 1 and 2 are static and are connected via a cable. Relay nodes and robots are connected if their distance is smaller than a certain threshold. Given all nodes with their current locations at time $t$, the service function $\lambda$ generates a graph $G$ shown in Fig. \ref{exp} based on the rules defined above. Since we have three types of robots, the spatial temporal trace for different nodes is given as follows. 
\begin{equation}
\begin{aligned}
    &\overrightarrow{x}(Robot,t)=[0 \quad 0]^T,\\
    &\overrightarrow{x}(Relay,t)=[0 \quad 1]^T,\\
    &\overrightarrow{x}(Base,t)=[1 \quad 1]^T.
\end{aligned}
\label{trace}
\end{equation}
\end{example}

Given the graph based spatial model and the definition of spatial temporal trace mentioned above, STREL is defined by extending STL with two novel spatial operators, ``reach" and ``escape". The syntax of STREL is given as follows \cite{bartocci2017monitoring}. 
\begin{definition}[STREL Syntax] 
The class of STREL formulas is defined recursively as
\begin{align*}
        \varphi::=&\mu~ |~\neg\varphi ~|~\varphi_1 \wedge \varphi_2 ~|\varphi_1 \vee \varphi_2  ~|~\Box_{[a,b]} \varphi~| ~ \varphi_1\sqcup_{[a,b]}\varphi_2  ~\\
        &|~ \varphi_1\mathcal{R}_d^f\varphi_2 ~|~ \mathcal{E}_d^f \varphi,
\end{align*}
where $\mu$ is an atomic predicate (AP), negation $\neg$, conjunction $\wedge$, and disjunction $\vee$ are the standard Boolean operators; $\Box_{[a,b]}$ is the ``always" operator and $\sqcup_{[a,b]}$ is the ``until" temporal operator, with $[a,b]$ as a real positive closed interval. Similarly as defined in STL, the ``eventually" operator $\Diamond$ can also be defined through ``always" and ``until" by setting $\Diamond_{[a,b]}\varphi=True\sqcup_{[a,b]}\varphi$. The spatial operators are the ``reach" $\mathcal{R}_d^f$ and the ``escape" $\mathcal{E}_d^f$ operators, where $f$ is a distance function and $d$ is a distance predicate. 
\end{definition}

Other spatial operators such as ``everywhere" $\boxbox_d^f\varphi$, ``somewhere" $\boxdot^f_d\varphi$,  and ``surround" $\varphi_1\varocircle_d^f\varphi_2$ can be derived from the ``reach" and ``escape" operators.
\begin{enumerate}
    \item Somewhere. $\boxdot^f_d\varphi=\emph{True}\mathcal{R}_d^f\varphi$ holds true for node $l$ and time $t$ if and only if there exists a node satisfying $\varphi$ and it is reachable from $l$ with distance constraints specified by $d$.
    \item Everywhere. $\boxbox_d^f\varphi$ is defined based on ``somewhere" operators by $\boxbox_d^f\varphi=\neg \boxdot^f_d\neg\varphi$. $\boxbox_d^f\varphi$ holds for node $l$ if and only if all nodes reachable from $l$ with distance bounded by $d$ satisfy $\varphi$.
    \item Surround. $\varphi_1\varocircle_d^f\varphi_2=\varphi_1\wedge\neg(\varphi_1\mathcal{R}_d^f\neg(\varphi_1\vee\varphi_2)\wedge\neg(\mathcal{E}^f_{\neg d}\varphi_1))$ holds for the node $l$ if and only if $l$ satisfies $\varphi_1$ and it cannot escape from the region satisfying $\varphi_1$ without passing any node satisfying $\varphi_2$ through a route with a length that satisfies constraints $d$.
\end{enumerate}

The semantics of STREL with respect to a discrete-time spatial temporal trace $\overrightarrow{x}(t_k,l)$ are introduced in Definition \ref{strelsem}, where $\overrightarrow{x}(t_k,l)\models\varphi$ denotes for which spatial temporal trace values and at what location and time index the formula $\varphi$ holds true. A signal interpretation function $\iota=[\iota_1,\iota_2,...,\iota_n]^T$ is needed and is defined as follows.
\begin{align*}
    \iota: AP \times D_x^n \rightarrow D^n,
\end{align*}
where $AP$ is a finite set containing all possible atomic propositions, and $D_x^n$ is the signal domain for spatial temporal trace $\overrightarrow{x}$. The interpretation function maps the atomic proposition and spatio-temporal trace into another domain such as Boolean or real value domain. 

\begin{definition}[STREL Semantics]\rm
The validity of an STR-
EL formula $\varphi$ with respect to signal $\overrightarrow{x}(t_k,l)$ at time $t_k$ and location $l$ is defined inductively as follows.
\begin{enumerate}
\item $\overrightarrow{x}(t_k,l)\models \mu$, if and only if $\iota_i(\mu,\overrightarrow{x}(t_k,l))>0,~\forall i\in [1,n]$;
\item $\overrightarrow{x}(t_k,l)\models \neg\varphi$, if and only if $\neg(\overrightarrow{x}(t_k,l))\models \varphi)$;
\item $\overrightarrow{x}(t_k,l)\models \varphi\land\psi$, if and only if $\overrightarrow{x}(t_k,l)\models \varphi$ and $\overrightarrow{x}(t_k,l)\models \psi$;
\item $\overrightarrow{x}(t_k,l)\models \varphi\lor\psi$, if and only if $\overrightarrow{x}(t_k,l)\models \varphi$ or $\overrightarrow{x}(t_k,l)\models \psi$;
\item $\overrightarrow{x}(t_k,l)\models \Box_{[a,b]}\varphi$, if and only if $\forall t_{k'}\in[t_k+a,t_k+b]$, $\overrightarrow{x}(t_k,l)\models \varphi$;
\item $\overrightarrow{x}(t_k,l)\models \varphi\sqcup_{[a,b]}\psi$, if and only if $\exists t_{k'}\in[t_k+a,t_k+b]$ such that $\overrightarrow{x}(t_k,l)\models \psi$ and $\forall t_{k''}\in[t_k,t_{k'}]$, $\overrightarrow{x}(t_k,l)\models \varphi$;
\item $\overrightarrow{x}(t_k,l)\models \varphi\mathcal{R}_d^f\psi$, if and only if $\exists \tau\in Routes(\lambda(t),l)$ $\exists~ l'\in\tau: (d_\tau^{f}(l,l')\vdash d)$ such that $\overrightarrow{x}(t_k,l')\models \psi$ and $\wedge_{j<\tau(l')}\overrightarrow{x}(t_k,\tau[j])\models \varphi$;
\item $\overrightarrow{x}(t_k,l)\models \mathcal{E}_d^f\varphi$, if and only if $\exists \tau\in Routes(\lambda(t),l)$ $\exists~l'\in\tau:(d_\tau^{f}(l,l')\vdash d)$ such that $\wedge_{i\leq\tau(l')}\overrightarrow{x}(t_k,\tau[i])\models \varphi$
\end{enumerate}
\label{strelsem}
\end{definition}
where $\lambda(t)$ is the service function, $Routes(\lambda(t),l)$ denotes an indexed sequence on the graph generated by the service function $\lambda(t)$ starting at node $l$, and $d_\tau^{f}(l,l')$ is the distance function between two nodes.

The intuitive idea of ``reach" operator $\varphi\mathcal{R}_d^f\psi$ is that there exists a route starting at $l$ with finite length $d_\tau^{f}(l,l')$ satisfying distance predicate $d$, which can reach a node satisfying $\psi$ and always satisfy $\varphi$ along the way. As for ``escape" operator $\mathcal{E}_d^f\varphi$, a node $l$ satisfies $\mathcal{E}_d^f\varphi$ if and only if there exists a route with finite length $d_\tau^{f}(l,l')$ satisfying distance predicate $d$ while all nodes on this route satisfy specification $\varphi$.

In addition to STREL syntax and semantics, the robustness of satisfaction of an STREL formula can be defined in a similar fashion as for STL by assigning a real-valued measure $m^\varphi$ for a spatial temporal trace $\overrightarrow{x}(t_k,l)$ at location $l$ and time $t_k$ such that $\overrightarrow{x}(t_k,l)\models\varphi$ if and only if $m^\varphi(\overrightarrow{x}(t_k,l))>0$.

\begin{definition}[Robustness of STREL]\label{def:robustSTREL}
The robustness of an STREL formula's satisfaction is inductively defined as follows.
\begin{enumerate}
  \item $\begin{aligned}[t]
     &m^{\mu}(\lambda,\overrightarrow{x},t,l)=\iota(\mu,\overrightarrow{x}(t,l));\\
  \end{aligned}$
  \item $\begin{aligned}[t]
     &m^{\neg\varphi}(\lambda,\overrightarrow{x},t,l)=\neg m(\lambda,\overrightarrow{x},\varphi,t,l);\\
  \end{aligned}$
  \item $\begin{aligned}[t]
    &m^{\varphi_1\wedge\varphi_2}(\lambda,\overrightarrow{x},t,l)\\
    &=\min(m^{\varphi_1}(\lambda,\overrightarrow{x},t,l),m^{\varphi_2}(\lambda,\overrightarrow{x},t,l));\\
  \end{aligned}$
    \item $\begin{aligned}[t]
    &m^{\varphi_1\vee\varphi_2}(\lambda,\overrightarrow{x},t,l)\\
    &=\max(m^{\varphi_1}(\lambda,\overrightarrow{x},t,l),m^{\varphi_2}(\lambda,\overrightarrow{x},t,l));\\
  \end{aligned}$
  \item $\begin{aligned}[t]
    &m^{\Box_{[a,b]}\varphi}(\lambda,\overrightarrow{x},t,l)=\min_{t'\in[t+a,t+b]}m^{\varphi}(\lambda,\overrightarrow{x},t',l);\\
  \end{aligned}$
  \item $\begin{aligned}[t]
    &m^{\varphi_1\sqcup_{[a,b]}\varphi_2}(\lambda,\overrightarrow{x},t,l)=\max_{t'\in[t+a,t+b]}(\\
    &\min(m^{\varphi_2}(\lambda,\overrightarrow{x},t,l)),\min_{t''\in[t,t']}(m^{\varphi_1}(\lambda,\overrightarrow{x},t'',l)));\\
  \end{aligned}$
  \item $\begin{aligned}[t]
     &m^{\varphi\mathcal{R}\psi}(\lambda,\overrightarrow{x},t,l)=\max_{\tau\in Routes(\lambda(t),l)}(\\
     &\max_{l'\in\tau:(d_\tau^{f}(l,l')\vdash d)}m^{\psi}(\lambda,\overrightarrow{x},t,l'),\\
     &\min_{j\leq \tau(l')}m^{\varphi}(\lambda,\overrightarrow{x},t,\tau[j]));\\
  \end{aligned}$
  \item $\begin{aligned}[t]
     &m^{\mathcal{E}_d^f\varphi}(\lambda,\overrightarrow{x},t,l)=\max_{\tau\in Routes(\lambda(t),l)}\\
     &\max_{l'\in\tau:(d_\tau^{f}(l,l')\vdash d)}\min_{i\leq \tau(l')}m^{\varphi}(\lambda,\overrightarrow{x},t,\tau[i]).
  \end{aligned}$
\end{enumerate}

\end{definition}

\section{Problem Statement}

\subsection{STL Motion Planning Specifications}
We now proceed to formulate a distributed communication-aware motion planning problem for a team of networked mobile robots. Let us consider the team of $P$ robots in a shared environment $\mathcal{X}$, each of which is governed by the discretized dynamics in (\ref{agentdynamic}). We assign a sequence of goal regions $\mathcal{X}_{i,q}^{goal}$ 
for the robot $\mathcal{R}_i$, where $i\in\mathcal{P}$ and $q\in Q_i$. $Q_i$ is the number of the sequence of goal regions for the robot $\mathcal{R}_i$ which are generated in Section V. Each region $\mathcal{X}_{i,q}^{goal}$ is characterized by a {\it polytope} \cite{kloetzer2008fully} in $\mathcal{X}_{free}$, i.e., there exists an integer $M_i\ge3$, a set of vectors $a_{i,q,j}\in \mathbb{R}^2$ and a set of scalars $b_{i,q,j}\in\mathbb{R}$, $j=1,2,\ldots, M_i$ for $\mathcal{R}_i$ such that
\begin{equation}
\mathcal{X}_{i,q}^{goal}=\{p\in\mathbb{R}^2|a_{i,q,j}^Tp+b_{i,q,j}\le 0, j=1,2,\ldots, M_i\}.
\end{equation}
In other words,
\begin{equation}
\begin{split}
\mathcal{X}_{i,q}^{goal}=\{&x\in\mathcal{X}|a_{i,q,j}^T[I_2\quad O_4]x+b_{i,q,j}\le 0,\\
&j=1,2,\ldots, M_i\},    
\end{split}
\end{equation}
where $I_n, O_n\in\mathbb{R}^{n\times n}$ denote the $n\times n$ identity and zero matrices, respectively.

Without loss of generality, we also assume that the region $\mathcal{X}_{obs}$ is  a polygonal subset of $\mathcal{X}$, i.e., there exists an integer $M_{obs}\ge 3$, a vector $a_{obs,j}\in \mathbb{R}^2$, and a scalar $b_{obs,j}\in\mathbb{R}$, with $j=1,2,\ldots, M_{obs}$, such that
\begin{equation}
\begin{split}
    \mathcal{X}_{obs}=\{&p\in\mathbb{R}^2|a_{obs,j}^Tp+b_{obs,j}\le 0,\\
    &j=1,2,\ldots, M_{obs}\}.
\end{split}
\end{equation}

We assume that all robots share a synchronized clock. The terminal time of multi-robot motion is upper-bounded by $t_f=T_f\Delta t$ with $T_f\in\mathbb{N}$, and the planning horizon is then given by $[0,T_f]$. Individual assignment is of practical importance, for instance, search and rescue missions or coverage tasks are often given to mobile robots individually. In this paper, local motion planning tasks for the robot $\mathcal{R}_i$ are summarized as follows. 

\begin{equation}
\begin{aligned}
&\forall i \in \mathcal{P},~\forall q\in Q_i,\\
&\varphi_{i,q}=\varphi_{i,p}\land\varphi_{i,s,col}\land\varphi_{i,s,obs},
\end{aligned}
\label{stl}
\end{equation}
where
\begin{enumerate}
\item the motion performance property
\begin{equation}
\varphi_{i,p}=\Diamond_{[0,T_f]} \bigwedge_{j=1}^{M_i}\left(a_{i,q,j}^T[I_2\quad O_4]x_i+b_{i,q,j}\le 0\right)
\label{stl1}
\end{equation}
requires that $\mathcal{R}_i$ enter the goal region within $T_f$ time steps;
\item the collision-avoidance safety property
\begin{equation}
\begin{aligned}
\varphi_{i,s,col} &= \Box_{[0,T_f]} \bigwedge_{j\in\mathcal{N}_i, j\ne i}\left[(|p_{i,1}-p_{j,1}|\ge d_1) \right.\\
&\left.\land(|p_{i,2}-p_{j,2}|\ge d_2)\right]
\end{aligned}
\end{equation}

ensures that $\mathcal{R}_i$ will never encounter collision with other robots. Here $d_1$ and $d_2$ are pre-defined safety distances between two robots in the two dimensions. $\mathcal{N}_i \subseteq \mathcal{P}$ denotes the set for the robot $\mathcal{R}_i$'s neighbor which will be described in Section \ref{MPC};
\item and the obstacle-avoidance safety property
\begin{equation}
\begin{aligned}
\varphi_{i,s,obs}=\Box_{[0,T_f]}\bigwedge_{j=1}^{M_{obs}}\left(a_{obs,j}^T[I_2\quad O_4]x_i+b_{obs,j}> 0\right)
\end{aligned}
\end{equation}
keeps the robot $\mathcal{R}_i$ from reaching any obstacles.
\end{enumerate}

\subsection{STREL Specifications of Communication}
In order to gather necessary information from the environment, such as the states of other robots and the positions of obstacles, a robot needs to establish communication links with its peers and with communication base stations. In this paper, we consider an indoor communication scenario shown in Fig. \ref{comm} where a team of robots needs to accomplish certain tasks while maintaining communication links among robots and base stations such that information can be shared within the team. Due to the path loss and shadowing effect caused by obstacles like walls in millimeter wave communication channel, communication base stations in each room can only reach robots within the same room. The base station located in the hall can only cover the blue area due to path loss. Robots may need  relay service from relay robots if they cannot reach any base station directly.  

\begin{figure}
 \centering
 \includegraphics[width=1\linewidth]{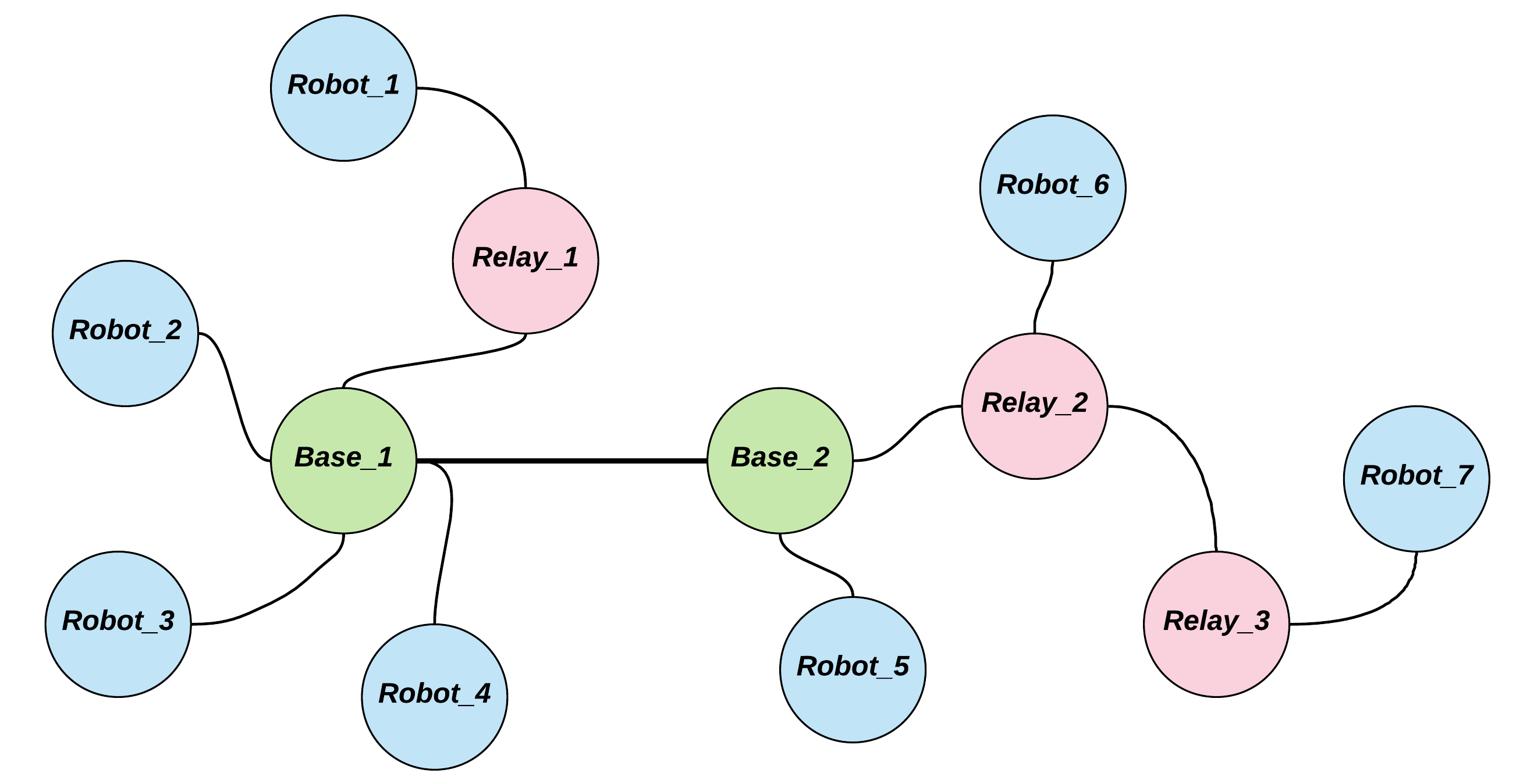}
 \caption{An example of STREL description of communication}
 \label{exp}
 \vspace{-5.5mm}
\end{figure}

To specify the spatial-temporal specifications over the mobile multi-robot system mentioned above, we formulate the STREL formula as follows.
\begin{equation}
    \psi=\Box_{[0,T_f]}(~\emph{robot}~\mathcal{R}_{d\leq 2}^{\emph{hops}}~\emph{base}~\bigwedge\boxbox_{d\leq 1}^\emph{hops}~\neg \emph{relay}).
    \label{strel}
\end{equation}
The formula (\ref{strel}) suggests that all robots reach a base station with no more than two hops and that relay nodes are not allowed to connect with each other in order to reduce communication delay caused by the relay. In Fig. \ref{exp}, all the robots except Robot\_7 satisfy $\psi$. Relay\_2 and Relay\_3 violate the everywhere specification as they are connected to each other. 

\subsection{MPC based Co-optimization Problem}
To pursue co-optimization for both motion planning and communication QoS for the networked robots, we use the following linear quadratic cost function $J_{i,1}$ to represent the energy consumption for the robot $\mathcal{R}_i$.

\begin{equation}
J_{i,1}=\sum_{k=k'}^{k'+H-1}\left(q^T|x_{i,t_k}|+r^T|u_{i,t_k}|\right)+h(k')\sum_{k=k'}^{k'+H-1}d_{i,k},
\label{J1}
\end{equation}
where $k'$ is the current time step, $H$ is the planning horizon, $q$ and $r$ are non-negative weighting vectors and $|.|$ denotes the element-wise absolute value. The second term is a time penalty multiplying a goal penalty such that each robot can move towards its goal. We define this goal penalty as $d_{i,k}=\|p_{i,k}^T-p_{i,goal}^T\|$ and the time penalty as $h(k)=\eta k^2$, where $p_{i,goal} \in \mathbb{R}^2$ denotes the geometric center of goal region $\mathcal{X}_{i,goal}$ for the robot $\mathcal{R}_i$, and $\eta$ is a parameter defined by the operator.

Regarding the communication QoS, we aim not only to satisfy the communication requirements specified by $\psi$ but also to increase the robustness of the communication network. According to Definition~\ref{def:robustSTREL}, we define the communication cost $J_{i,2}$ for the robot $\mathcal{R}_i$ as follows.

\begin{equation}
J_{i,2}=-m^{\psi}(\lambda,\overrightarrow{x},t,l).
\label{J2}
\end{equation}

Similar to the robustness of STL, the STREL robustness can be computed recursively based on the structure of the formula with additional constraints. Since $\min$ and $\max$ operators can be encoded as MILP with additional binary variables \cite{raman2014model}, the communication cost $J_{i,2}$ can be encoded as MILP.

Based on the aforementioned preliminaries and cost functions, we formulate the distributed communication-aware motion planning problem for the networked robots under the STL-STREL specifications from an MPC perspective.
\begin{problem}\label{prob:motion planning}
	Let us consider a networked mobile robots system with $P$ robots whose dynamics are given by (\ref{agentdynamic}) with initial states $x_{i,0}$, 
	 and a planning horizon $H>0$. Motion planning and communication requirements are specified by a local STL formula $\varphi_i$ in (\ref{stl}) and an STREL formula $\psi$ in (\ref{strel}), respectively. We aim to find the local control input $u_i(t_k)$, which is the first element of the sequence ${\bf u}_i^{H}=u_i(t_k)u_i(t_{k+1})\ldots u_i(t_{k+H-1})$, for the robot $\mathcal{R}_i$ $(i\in \mathcal{P})$ that solves the following co-optimization problem:
	\begin{equation}
	\underset{{\bf u}_i^{H}, i\in\mathcal{P}}{\min}\quad J_i\left({\bf x}_i(x_{i,0},{\bf u}_i^{H})\right)=\alpha J_{i,1}+(1-\alpha)J_{i,2},
	\label{J}
	\end{equation}
	\begin{alignat*}{3}
	\mbox{s.t.}\quad &\forall i\in\mathcal{P}, \quad \forall q \in Q_i, \\
	&x_i(t_{k+1})=A_{i,d}x_i(t_k)+B_{i,d}u_i(t_k),\\
	&{\bf x}_i(x_{i,k},{\bf u}_i^{H}) \models \varphi_{i,q},\\
	&(\overrightarrow{x},t_k,l)\models \psi,\\
	&u_i\in\mathcal{U}=[-u_{max}, u_{max}]\times[-u_{max}, u_{max}], \\
	&||v_i||<v_{max},\\
	& \omega_i=\frac{||u_i||}{m_i||v_i||}\leq\frac{u_{max}}{m_iv_{max}},
	\end{alignat*}
where $\alpha \in [0,1]$ is a design parameter determined by the operator, $u_{max}$ and $v_{max}$ are constants bounding $u_i$ and $v_i$, $w_i$ is the turning rate, and $m_i$ denotes the mass of robot $\mathcal{R}_i$.	
\end{problem}

\section{MILP Encoding of Communication-aware Motion Planning}

\subsection{MILP Encoding of Robot Dynamics}
For the sake of brevity, we replace $t_k$ with $t$ in this section and denote the state and control inputs of the robot $\mathcal{R}_i$ at time step $t$ as $x_{it}$ and $u_{it}$, respectively. To encode the motion planning cost (\ref{J1}) as linear programming, we employ Manhattan distance for $d_{i,k}$ and introduce slack vectors $\alpha_{it}$, $\beta_{it}$, $\gamma_{it}$ and additional constraints \cite{athans2013optimal} such that $J_{i,1}$ can be transformed into the following linear cost function.
\begin{equation}
J_{i,1}=\sum_{t=k'}^{k'+H-1}({q}^T{\alpha_{it}+{r}^T\beta_{it}})+h(k')\sum_{t=k'}^{k'+H-1}\sum_{k=1}^{2}\gamma_{itk},
\end{equation}

\begin{equation}
\begin{aligned}
&\text{s.t.} & \forall t\in [k',k'+&H-1], \forall j\in [1,4], \forall k\in [1, 2], \\
& \text{}
& x_{itj}&\leq \alpha_{itj}, -x_{itj}\leq \alpha_{itj}, \\
& \text{and}
& u_{itk}&\leq \beta_{itk}, -u_{itk}\leq \beta_{itk},\\
& \text{and}
& x_{itk}-p_{i,goal,k}&\leq \gamma_{itk}, -x_{itk}+p_{i,goal,k}\leq \gamma_{itk} \\
& \text{and}
& x_i(t+1)&=A_{i,d}x_i(t)+B_{i,d}u_i(t).
\end{aligned}
\end{equation}

To transfer the nonlinear velocity constraints, we use the method in \cite{richards2002aircraft} which is introducing an arbitrary number $L$ of linear constraints such that the 2-D velocities can be approximated by a regular L-sided polygon.
\begin{equation}
\begin{split}
\forall l\in [1,L], i\in[1,P],t\in[k',k'+H-1],\\
v_{it1}sin(\frac{2\pi l}{L})+v_{it2}cos(\frac{2\pi l}{L})\leq v_{max}.
\end{split}
\end{equation}

\subsection{Boolean Encoding of STL Constraints}

For the MILP encoding of STL specifications in (\ref{stl}), we denote three Boolean variables $z_{t}^{\varphi_{i,p}}$, $z_{t}^{\varphi_{i,s,col}}$, and $z_{t}^{\varphi_{i,s,obs}}$ whose value depends on the satisfaction of $\varphi_{i,p}$ , $\varphi_{i,s,col}$, and $\varphi_{i,s,obs}$, respectively \cite{raman2014model,liuacc}. The satisfaction of $\varphi_i$ at time step $t$ can therefore be represented by the Boolean variable $z_{t}^{\varphi_i}$ which is determined by $z_{t}^{\varphi_{i,p}}$, $z_{t}^{\varphi_{i,s,col}}$ and $z_{t}^{\varphi_{i,s,obs}}$ as follows.

\begin{equation}
z_{t}^{\varphi_i}=z_{t}^{\varphi_{i,p}} \land z_{t}^{\varphi_{i,s,col}} \land z_{t}^{\varphi_{i,s,obs}},
\end{equation}
with
\begin{equation}
\begin{split}
\forall i\in\mathcal{P}: \\
&z_{t}^{\varphi_i}\le z_{t}^{\varphi_{i,p}}, z_{t}^{\varphi_i}\le z_{t}^{\varphi_{i,s,col}}, z_{t}^{\varphi_i}\le z_{t}^{\varphi_{i,s,obs}},\\
&z_{t}^{\varphi_i}\ge z_{t}^{\varphi_{i,p}}+z_{t}^{\varphi_{i,s,col}}+z_{t}^{\varphi_{i,s,obs}}-2,
\end{split}
\end{equation}
where $z_{t}^{\varphi_{i,p}}$, $z_{t}^{\varphi_{i,s,col}}$, and $z_{t}^{\varphi_{i,s,obs}}$ are 1 if and only if their corresponding specifications are satisfied.

\subsubsection{Boolean Encoding of $\varphi_{i,p}$}
Recall (\ref{stl1}). For robot $\mathcal{R}_i$, we re-write $\varphi_{i,p}$ as follows.

\begin{equation}
\varphi_{i,p}=\Diamond_{[0,T_f]} \bigwedge_{j=1}^{M_i}\neg\pi^{\mu_{i,j}},
\end{equation}
where $\pi^{\mu_{i,j}}$ is a predicate for $i\in\mathcal{P}$ and $j\in[1, M_i]$, with
\begin{equation}
\mu_{i,j}(x_i)=-a_{i,j}^T[I_2\quad O_4]x_i-b_{i,j}
\label{con4}
\end{equation}
being the function whose sign determines whether $\pi^{\mu_{i,j}}$ is true or not. In this case, we associate a Boolean variable $z_{t}^{\pi^{\mu_{i,j}}}$ with $\pi^{\mu_{i,j}}$ for $t\in[0,T_f]$ and deploy the following constraints \cite{raman2014model} to enforce $z_{t}^{\pi^{\mu_{i,j}}}=1$ if and only if $\mu_{i,j}(x_i(t))>0$.
\begin{equation}
\begin{split}
\mu_{i,j}(x_i(t))\le Mz_{t}^{\pi^{\mu_{i,j}}}-\varepsilon, \\
-\mu_{i,j}(x_i(t))\le M(1-z_{t}^{\pi^{\mu_{i,j}}})-\varepsilon,
\end{split}
\label{con3}
\end{equation}
where $M$ is selected to be a sufficiently large positive real number, and $\varepsilon$ is a sufficiently small positive real number such that $z_{t}^{\pi^{\mu_{i,j}}}=1$ if and only if $\mu_{i,j}(x_i(t))>0$.

For the negation $\neg\pi^{\mu_{i,j}}$ of the predicate $\pi^{\mu_{i,j}}$, it is clear that the corresponding Boolean variable is
\begin{equation}
z_{t}^{\neg\pi^{\mu_{i,j}}}=1-z_{t}^{\pi^{\mu_{i,j}}},
\end{equation}
whose corresponding MILP constraints can be re-written from (\ref{con3}) as follows.
\begin{equation}
\begin{split}
\mu_{i,j}(x_i(t))\le M(1-z_{t}^{\neg\pi^{\mu_{i,j}}})-\varepsilon, \\
-\mu_{i,j}(x_i(t))\le Mz_{t}^{\neg\pi^{\mu_{i,j}}}-\varepsilon,
\end{split}
\end{equation}
where $\mu_{i,j}(x_i(t))$ is given by (\ref{con4}).

Let $\psi_i=\bigwedge_{j=1}^{M_i}\neg\pi^{\mu_{i,j}}$. Then, the corresponding Boolean variable for $\psi_i$, namely $z_{t}^{\psi_i}$, is given by
\begin{equation}
z_{t}^{\psi_i}=\bigwedge_{j=1}^{M_i} z_{t}^{\neg\pi^{\mu_{i,j}}},
\end{equation}
with the following extra constraints \cite{raman2014model}
\begin{equation}
\begin{split}
\forall &j\in[1, M_i], \\
& z_{t}^{\psi_i} \le z_{t}^{\neg\pi^{\mu_{i,j}}}, \\
&z_{t}^{\psi_i} \ge \sum_{j=1}^{M_i} z_{t}^{\neg\pi^{\mu_{i,j}}}-M_i+1.
\end{split}
\end{equation}

Finally, we can formally write $z_{t}^{\varphi_{i,p}}$ by incorporating $z_{t}^{\psi_i}$ with the $\Diamond$ operator. Towards this end, we have
\begin{equation}
z_{t}^{\varphi_{i,p}}=\bigvee_{t'=t}^{T_f} z_{t'}^{\psi_i},
\end{equation}
which implies that the performance specification $\varphi_{i,p}$ will be satisfied at some time step within the interval $[t,T_f]$.

\subsubsection{Boolean Encoding of $\varphi_{i,s,col}$ and $\varphi_{i,s,obs}$}
First, similar to the encoding process of $\psi_i=\bigwedge_{j=1}^{M_i}\neg\pi^{\mu_{i,j}}$, we consider
\begin{equation}
\psi_i^{obs}=\bigwedge_{j=1}^{M_i}\pi^{\mu^{obs}_{i,j}},    
\end{equation}
in which $\pi^{\mu^{obs}_{i,j}}$ is a predicate with the corresponding function $$\mu^{obs}_{i,j}(x_i)=a_{obs,j}^T[I_2\quad O_4]x_i+b_{obs,j}$$
for $j=[1,M_{obs}]$. Boolean variables $z_{t}^{\pi^{\mu^{obs}_{i,j}}}$, $j\in[1,M_{obs}]$, are added such that $z_{t}^{\pi^{\mu^{obs}_{i,j}}}=1$ if and only if $\mu^{obs}_{i,j}(x_i(t))>0$, which results in the following Boolean encoding

\begin{equation}
z_{t}^{\psi^{obs}_i}=\bigwedge_{j=1}^{M_{obs}} z_{t}^{\pi^{\mu^{obs}_{i,j}}},
\end{equation}
with MILP constraints

\begin{equation}
\begin{split}
\forall &j\in[1, M_{obs}],\\
& z_{t}^{\psi^{obs}_i} \le z_{t}^{\pi^{\mu^{obs}_{i,j}}}, \\
&z_{t}^{\psi^{obs}_i} \ge \sum_{j=1}^{M_{obs}} z_{t}^{\pi^{\mu^{obs}_{i,j}}}-M_{obs}+1, \\
&\mu^{obs}_{i,j}(x_i(t))\le Mz_{t}^{\pi^{\mu_{i,j}}}-\varepsilon, \\
&-\mu^{obs}_{i,j}(x_i(t))\le M(1-z_{t}^{\pi^{\mu_{i,j}}})-\varepsilon,
\end{split}
\label{con1}
\end{equation}
where $ M$ and $\varepsilon$ are appropriately-selected positive real numbers. Therefore, the Boolean variable for the satisfaction of $\varphi_{i,s,obs}$ is given by
\begin{equation}
\begin{split}
z_{t}^{\varphi_{i,s,obs}}&=z_{t}^{\Box_{[0,T_f]}\psi^{obs}_i} =\bigwedge_{t=0}^{T_f} z_{t}^{\psi^{obs}_i}
\end{split}
\label{con2}
\end{equation}
with constraints (\ref{con1}).

On the other hand, the collision-avoidance safety property $\varphi_{i,s,col}$ requires that any robot travel outside a rectangle with length $2d_1$ and width $2d_2$ centered at the robot $\mathcal{R}_i$'s position. Since the control $u_i$ for each $i\in\mathcal{P}$ is local for each robot, state information of other robots is obtained through a inter-robot communication channel, whose QoS is optimized in the next section. For any $j\in\mathcal{P}\setminus\{i\}$, we have four predicates, $\pi^{\mu^1_{i,j}}$, $\pi^{\mu^2_{i,j}}$, $\pi^{\mu^3_{i,j}}$, and $\pi^{\mu^4_{i,j}}$ whose respective corresponding functions are listed as follows.
\begin{equation}
\begin{split}
\mu^1_{i,j}(x_i)&=[1~ 0~ 0~ 0~0~0]x_i-[1~ 0~ 0~ 0~0~0]x_j-d_1, \\
\mu^2_{i,j}(x_i)&=[1~ 0~ 0~ 0~0~0]x_j-[1~ 0~ 0~ 0~0~0]x_i-d_1, \\
\mu^3_{i,j}(x_i)&=[0~ 1~ 0~0~0~0]x_i-[0~ 1~ 0~0~0~0]x_j-d_2, \\
\mu^4_{i,j}(x_i)&=[0~ 1~ 0~0~0~0]x_j-[0~ 1~ 0~0~0~0]x_i-d_2.
\end{split}
\end{equation}

The Boolean encoding of $\pi^{\mu^1_{i,j}}$, $\pi^{\mu^2_{i,j}}$, $\pi^{\mu^3_{i,j}}$, and $\pi^{\mu^4_{i,j}}$, leads to the introduction of four Boolean variables $z_{t}^{\pi^{\mu^1_{i,j}}}$, $z_{t}^{\pi^{\mu^2_{i,j}}}$, $z_{t}^{\pi^{\mu^3_{i,j}}}$, and $z_{t}^{\pi^{\mu^4_{i,j}}}$ with the following MILP constraints similar in structure to those of (\ref{con3}).

\begin{equation}
\begin{split}
\forall j\in\mathcal{P}& \setminus\{i\}, m\in[1,4], \\
&\mu^m_{i,j}(x_i(t))\le Mz_{t}^{\pi^{\mu^m_{i,j}}}-\varepsilon, \\
&-\mu^m_{i,j}(x_i(t))\le M(1-z_{t}^{\pi^{\mu^m_{i,j}}})-\varepsilon.
\end{split}
\end{equation}

Therefore, the Boolean variable for the satisfaction of $\varphi_{i,s,col,j}=\bigwedge_{m=1}^4 \pi^{\mu^m_{i,j}}$ for $j\in\mathcal{P}\setminus\{i\}$ is given by $$z_{t}^{\varphi_{i,s,col,j}}=\bigwedge_{m=1}^4 z_{t}^{\pi^{\mu^m_{i,j}}}$$
with
\begin{equation}
\begin{split}
z_{t}^{\varphi_{i,s,col,j}}&\le z_{t}^{\pi^{\mu^m_{i,j}}}, \\
z_{t}^{\varphi_{i,s,col,j}} &\ge \sum_{m=1}^{4} z_{t}^{\pi^{\mu^{m}_{i,j}}}-3.
\end{split}
\end{equation}
Thus, for the Boolean variable
$$z_{t}^{\bigwedge_{j\ne i}\varphi_{i,s,col,j}}=\bigwedge_{j\in\mathcal{P}\setminus\{i\}} z_{t}^{\varphi_{i,s,col,j}},$$
one can write
\begin{equation}
\begin{split}
\forall i\in\mathcal{P},~ &j\in\mathcal{P}\setminus\{i\}, \\
&z_{t}^{\bigwedge_{j\ne i}\varphi_{i,s,col,j}}\le z_{t_k}^{\varphi_{i,s,col,j}},\\
&z_{t}^{\bigwedge_{j\ne i}\varphi_{i,s,col,j}} \ge \sum_{j\in\mathcal{P}\setminus\{i\}} z_{t}^{\varphi_{i,s,col,j}}-(P-1)+1.
\end{split}
\end{equation}

It follows that the Boolean variable $z_{t}^{\varphi_{i,s,col}}$ is given by
\begin{equation}
z_{t}^{\varphi_{i,s,col}}=\bigwedge_{t=0}^{T_f} z_{t}^{\bigwedge_{j\ne i}\varphi_{i,s,col,j}}.
\end{equation}

\subsection{Boolean Encoding of STREL Constraints}

Similar to the MILP encoding of STL formulas, we encode the given STREL specifications $\psi$ in (\ref{strel}) by introducing a Boolean variable $z^\psi_{t,l}$ whose value depends on the satisfaction of $\psi$ at time $t$ and location $l$. $z^\psi_{t,l}=1$ as long as the corresponding STREL formula $\psi$ holds at time $t$ and location $l$. In this subsection, we first give a general procedure of encoding any STREL formulas as MILP. Then we illustrate the procedure by encoding the STREL formula in our problem. Based on the syntax of STREL, the Boolean encoding can be achieved through the following recursive process. 

\textbf{Predicates}: Given predicate $\mu$ at time $t$ and location $l$, we denote a Boolean variable $z^\mu_{t,l}$ whose truth value depends on the satisfaction of $\varphi=\mu$ at time $t$ and location $l$. To make sure $z^\mu_{t,l}=1$ if and only if $\varphi=\mu$, according to the STREL semantic definition, $z^\mu_{t,l}$ needs to satisfy the following constraints.
\begin{equation}
\begin{split}
\iota_i(\mu,\overrightarrow{x}(t,l))&\leq Mz^{\mu,i}_{t,l},\\
-\iota_i(\mu,\overrightarrow{x}(t,l))&\leq M(1-z^{\mu,i}_{t,l})-\epsilon,\\
z^{\mu}_{t,l}=\bigwedge_{i=1}^{n}z^{\mu,i}_{t,l},
\end{split}   
\end{equation}
where $M$ is a sufficiently large positive number such that $M>max(\iota(\mu,\overrightarrow{x}(t,l)))$ for all $t$ and $l$, and $\epsilon$ is a sufficiently small positive number close to 0. For any linear $\iota(\mu,\overrightarrow{x}(t,l))$, the given predicate $\mu$ can be encoded as MILP. 

\textbf{Negation}: Given STREL formula $\psi=\neg\varphi$ with negation operator, the corresponding Boolean variable $z^\psi_{t,l}$ should satisfy the following constraint.
\begin{align}
    z^\psi_{t,l}=1-z^\varphi_{t,l}.
\end{align}

\textbf{Conjunction}: For STREL formula $\varphi=\wedge_{i=1}^n\varphi_i$ containing conjunction operators, the satisfaction of the STREL formula depends on the Boolean variable $z^\varphi_{t,l}$, with the following linear constraints.
\begin{equation}
    \begin{split}
    z^\varphi_{t,l}&\leq z^{\varphi_i}_{t,l}, i=1,...,n, \\
    z^\varphi_{t,l}&\geq \sum_{i=1}^n z^{\varphi_i}_{t,l}-n+1.
    \end{split}
\end{equation}

\textbf{Disjunction}: STREL formula $\varphi=\vee_{i=1}^n\varphi_i$ with disjunction operators can be encoded as a Boolean variable $z^\varphi_{t,l}$ using a similar approach mentioned above.
\begin{equation}
    \begin{split}
    z^\varphi_{t,l}&\geq z^{\varphi_i}_{t,l}, i=1,...,n, \\
    z^\varphi_{t,l}&\leq \sum_{i=1}^n z^{\varphi_i}_{t,l}.
    \end{split}
\end{equation}

\textbf{Always}: Given STREL formula $\psi=\square_{[a,b]}\varphi$ with ``always" operators, the STREL formula $\psi$ holds true if and only if the corresponding Boolean variable $z^\psi_{t,l}$ equals to 1 subjecting to the following constraint.
\begin{align}
    z^\psi_{t,l}=\bigwedge_{i=a}^bz^{\varphi_i}.
\end{align}

\textbf{Eventually}: Given STREL formula $\psi=\diamond _{[a,b]}\varphi$ with ``eventually" operators, the STREL formula $\psi$ holds true if and only if the corresponding Boolean variable $z^\psi_{t,l}$ equals to one subjecting to the following constraint.
\begin{align}
    z^\psi_{t,l}=\bigvee_{i=a}^bz^{\varphi_i}.
\end{align}

\textbf{Until}: The bounded ``until" operator $\psi=\varphi_1\sqcup_{[a,b]}\varphi_2$ can be obtained through an unbounded ``until" operator as follows \cite{donze2013efficient}.
\begin{align*}
    \varphi_1\sqcup_{[a,b]}\varphi_2=\Box_{[0,a]}\varphi_1\bigwedge\Diamond_{[a,b]}\varphi_2\Diamond_{[a,a]}(\varphi_1\sqcup\varphi_2),
\end{align*}
where the unbounded ``until" operators can be encoded recursively as follows.
\begin{align*}
    &z_{t,l}^{\varphi_1\sqcup\varphi_2}=z_{t,l}^{\varphi_2}\bigvee(z_{t,l}^{\varphi_1}\wedge z_{t+1,l}^{\varphi_1\sqcup\varphi_2}), \forall t\in [1,...,N-1],\\
    &z_{N,l}^{\varphi_1\sqcup\varphi_2}=z_{N,l}^{\varphi_2}.
\end{align*}
Thus, the bounded ``until" operator can be encoded as
\begin{align}
    z_{t,l}^{\varphi_1\sqcup_{[a,b]}\varphi_2}=z_{t,l}^{\Box_{[0,a]}\varphi_1}\bigwedge z_{t,l}^{\Diamond_{[a,b]}\varphi_2}\bigwedge z_{t,l}^{\Diamond_{[a,a]}(\varphi_1\sqcup\varphi_2)}.
\end{align}

\textbf{Reach}: For the STREL formula $\psi=\varphi_1\mathcal{R}_d^f\varphi_2$ with a ``reach" spatial operator, we encode the ``reach" operator in a recursive fashion as well.
\begin{equation}
    \begin{split}
        &z_{t,l}^{\varphi_1\mathcal{R}_d^f\varphi_2}=\bigvee_{l'\in d_{\textbf{1}}}z_{t,l'}^{\varphi_1\mathcal{R}_{d-1}^f\varphi_2}\bigwedge(z_{t,l}^{\varphi_1}\vee z_{t,l}^{\varphi_2})\bigvee z_{t,l}^{\varphi_2},\\
    &z_{t,l}^{\varphi_1\mathcal{R}_{0}^f\varphi_2}=z_{t,l}^{\varphi_2},
    \end{split}
\end{equation}
where $l'\in d_{\textbf{1}}$ denotes the nodes that are one hop away to the node $l$. The intuitive idea of encoding the ``reach" operator this way is $\varphi_1\mathcal{R}_{d}^f\varphi_2$ holds for node $l$ if and only if there exists at least one adjacent node $l'$ which can reach $\varphi_2$ with one step less than $l$, while satisfying $\varphi_1$ along the way, or node $l$ itself already satisfies $\varphi_2$.

\textbf{Escape}: For STREL formulas $\mathcal{E}_d^f \varphi$ with an ``escape" spatial operator, we encode the ``escape" operator recursively as follows since $\mathcal{E}_d^f \varphi$ holds for node $l$ if and only if there exists at least one adjacent node $l'$, which can escape $\varphi$ with one step less than $l$, and the node itself also satisfies $\varphi$.
\begin{equation}
    \begin{split}
    &z_{t,l}^{\mathcal{E}_d^f\varphi}=\bigvee_{l'\in d_{\textbf{1}}}z_{t,l'}^{\mathcal{E}_{d-1}^f \varphi}\bigwedge z_{t,l}^{\varphi},\\
    &z_{t,l}^{\mathcal{E}_0^\varphi}=z_{t,l}^\varphi.
    \end{split}
\end{equation}

\begin{example}
Let us consider the STREL formula used in this paper and encode it as MILP using the method proposed above. 

\begin{equation}
\begin{aligned}
    &\psi=\Box_{[0,T_f]}\varphi,~ \varphi=\phi_1\wedge\phi_2, \\
    &\phi_1=\emph{robot}~\mathcal{R}_{d\leq 2}^{\emph{hops}}~\emph{base},~\phi_2=\boxbox_{d\leq 1}^\emph{hops}~\neg \emph{relay}.
    \end{aligned}
\end{equation}

Define Boolean variables $z^{\psi}_{t,l}$, $z^{\varphi}_{t,l}$, $z^{\phi_1}_{t,l}$, and $z^{\phi_2}_{t,l}$ whose values are 1 if and only if their corresponding STREL formulas hold for node $l$ at time $t$. Following the Boolean encoding procedure of STREL, these variables should satisfy the following Boolean encoding
\begin{equation}
    \begin{aligned}
    &z_{t,l}^{\psi}=\bigwedge_{t=0}^{T_f}z_{t,l}^{\varphi},\\
    &z_{t,l}^{\varphi}=z_{t,l}^{\phi_1}\wedge z_{t,l}^{\phi_2},
    \end{aligned}
\end{equation}
with MILP constraints
\begin{equation}
    \begin{aligned}
    &z_{t,l}^{\psi}\leq z_{t,l}^{\varphi},~ t=0,...,T_f,\\
    &z_{t,l}^{\psi}\geq \sum_{t=0}^T z_{t,l}^{\varphi}-T_f,\\
    &z_{t,l}^{\varphi_i}\leq z_{t,l}^{\phi_j},~ j=1,2,\\
    &z_{t,l}^{\varphi_i} \geq z_{t,l}^{\phi_1}+z_{t,l}^{\phi_2}-1.
    \end{aligned}
\end{equation}

The value of Boolean variable $z_{t,l}^{\phi_1}$ can be determined through the following procedure. 

\begin{align*}
    &z_{t,l}^{\emph{robot}~\mathcal{R}_{d\leq 2}^{\emph{hops}}~\emph{base}}=\bigvee_{l'\in d_{\textbf{1}}}z_{t,l'}^{\emph{robot}~\mathcal{R}_{d\leq 1}^{\emph{hops}}~\emph{base}}\bigwedge(z_{t,l}^{\emph{robot}}\vee z_{t,l}^{\emph{base}})\bigvee z_{t,l}^{\emph{base}},\\
    &z_{t,l'}^{\emph{robot}~\mathcal{R}_{d\leq 1}^{\emph{hops}}~\emph{base}}=\bigvee_{l''\in d_{\textbf{1}}}z_{t,l''}^{\emph{robot}~\mathcal{R}_{0}^{\emph{hops}}~\emph{base}}\bigwedge(z_{t,l'}^{\emph{robot}}\vee z_{t,l'}^{\emph{base}})\bigvee z_{t,l'}^{\emph{base}},\\
    &z_{t,l''}^{\emph{robot}~\mathcal{R}_{0}^{\emph{hops}}~\emph{base}}=z_{t,l''}^{\emph{base}}.
\end{align*}
$z_{t,l}^{\emph{robot}}$ is determined through the following equations, and $z_{t,l}^{\emph{base}}$ can be determined by the same procedure. 

\begin{align*}
    &z_{t,l}^{\emph{robot}}=\bigwedge_{i=1}^{2}z_{t,l}^{\emph{robot},i},\\
    &\forall i=1,2,\\
    &\iota_i(\emph{robot},\overrightarrow{x}(t,l))\leq Mz_{t,l}^{\emph{robot},i},\\
    -&\iota_i(\emph{robot},\overrightarrow{x}(t,l))\leq M(1-z_{t,l}^{\emph{robot},i})-\epsilon,
\end{align*}
where the interpretation function $\iota(\mu,\overrightarrow{x}(t,l))$ in this paper is chosen as a Boolean function with a set of linear constraints and defined as follows

\begin{equation}
\begin{aligned}
    &\iota(\mu,\overrightarrow{x}(t,l))=\iota(\begin{bmatrix}
\mu_1\\\mu_2
\end{bmatrix},\overrightarrow{x}(t,l))\\
&=\iota(\begin{bmatrix}
\mu_1\\\mu_2
\end{bmatrix},\begin{bmatrix}
a_1\\a_2
\end{bmatrix})=\begin{bmatrix}
1-|\mu_1-a_1|\\1-|\mu_2-a_2|
\end{bmatrix}=\begin{bmatrix}
\iota_1\\\iota_2
\end{bmatrix},\\
\end{aligned}
\label{iota}
\end{equation}
where $\mu_1$, $\mu_2$, $a_1$, and $a_2$ are all binary constants. Let us use Fig. \ref{exp} as an example to illustrate how the interpretation function $\iota(\mu,\overrightarrow{x}(t,l))$ works. For Robot\_1 in Fig. \ref{exp}, according to (\ref{trace}) we have $\iota(\emph{robot},\overrightarrow{x}(t,l))=\iota(\begin{bmatrix}
0\\0
\end{bmatrix},\begin{bmatrix}
0\\0
\end{bmatrix})$. From (\ref{iota}), we can conclude $\iota_1=1$ and $\iota_2=1$. From the definition of $\iota(\mu,\overrightarrow{x}(t,l))$ and the example above, we can see that $\iota_1=\iota_2=1$ if and only if node $l$ is the same type of robot specified in $\mu$.

\end{example}

\section{Distributed MPC Synthesis with a Guarantee of Completeness}\label{MPC}
\subsection{Distributed MPC with a Guarantee of Completeness}
To reduce the size of the co-optimization problem, we apply a distributed MPC framework in our previous work \cite{liu2017distributed}. Instead of solving optimization problems for the whole time horizon $T_f$, MPC solves problems within a finite horizon $H<T_f$ starting from current states, and provides finite control inputs $u^H_k$. Only the first control input will be implemented, and the states of robots will be sampled again in the next time step. 
It reduces the size of the problem by only considering $H$ steps ahead. For distributed MPC, each robot only considers its neighbors and solves its own MPC-based optimization problem which further reduces the size of the problem  \cite{kuwata2011cooperative,liu2017distributed}. Therefore, the distributed MPC framework makes the optimization problem much smaller than the centralized version. 

\begin{figure}[h]
	\centering
	\includegraphics[width=0.9\linewidth]{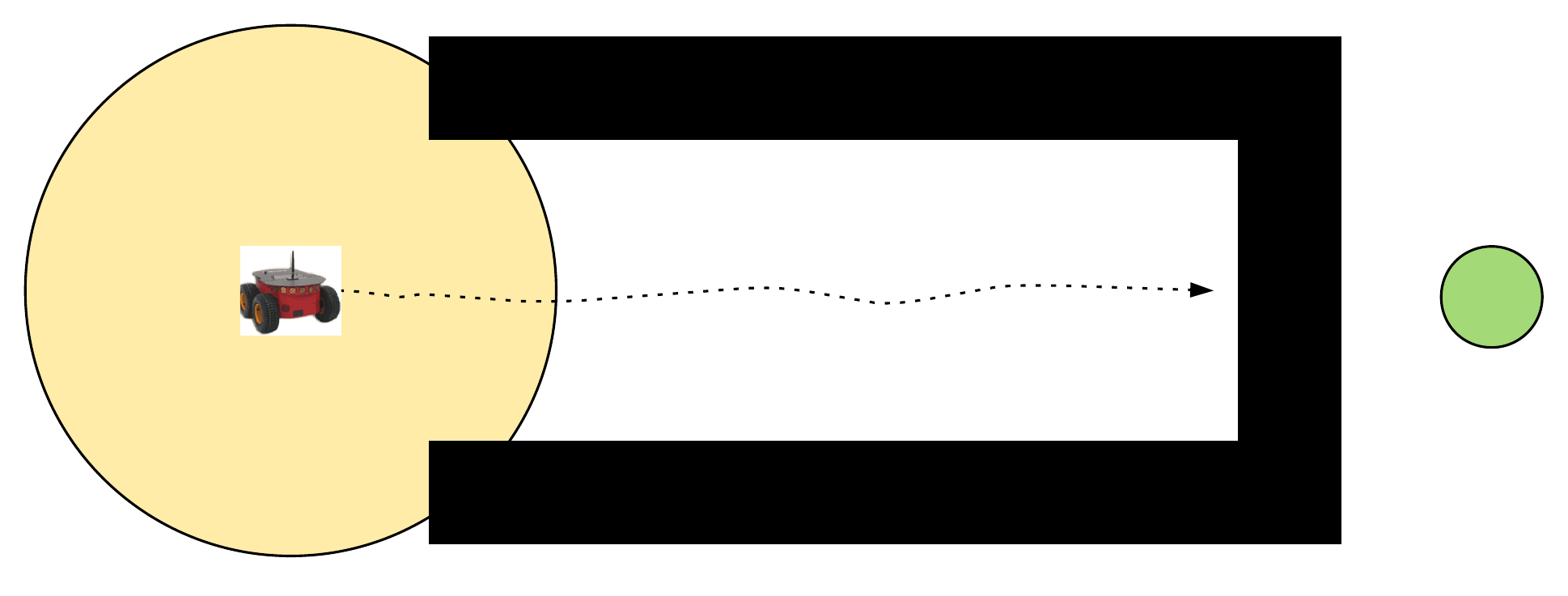}
	\caption{A robot with a limited planning range may get stuck at a dead end.}
	\label{deadend}
	\vspace{-3.5mm}
\end{figure}

Although introducing a distributed MPC framework can significantly reduce the computational complexity of communication-aware motion planning for multi-robot systems \cite{liu2017distributed}, completeness issues may arise during the implementation. Due to the limited and finite planning horizon, robots driven by the desire to minimize terminal cost in their motion planning cost function may get stuck in a local dead end even though a feasible global  solution exists. Fig. \ref{deadend} illustrates this scenario, where a robot with a limited planning horizon try to reach the target area marked by green. Since the robot attempts to minimize its motion planning cost in each planning period, it will drive into the dead end and be unable to escape. 

Several existing studies have addressed the completeness issue of MPC in motion planning \cite{liu2016high,watterson2015safe}. A two-layer MPC framework with different planning horizons was proposed by Watterson and Kumar \cite{watterson2015safe}, where a short-range receding horizon control layer was responsible for efficient local motion planning. Also, a long-range receding horizon control layer was invoked when the former layer failed to proceed, and planed a trajectory based on a graph search. The switch between short range and long range planning horizon enables efficient motion planning without sacrificing completeness. 

\subsection{The Two-layer Hierarchical Framework}
Based on ideas from these methods, we propose a two-layer hierarchical MPC framework where a low layer with a short planning horizon is responsible for the local communication-aware motion planning. A high layer is built on top of the local layer to guarantee the completeness of our proposed MPC framework. This high layer will generate a sequence of  waypoints to guide the low layer such that robots with a limited planning horizon will be able to escape a dead end. To generate the high layer waypoints, we first construct a global map based on the information of the indoor environment. The global map $\mathcal{G}=(\mathcal{V},\mathcal{W})$ is built upon a graph-based model where nodes represent partitions of the environment, and weights model the connection among the partitions. The partitions are constructed by 2-D Delaunay triangulation. 2-D Delaunay triangulation for a given set $\mathcal{O}$ of points partitions the environment without overlap using triangles where all vertices come from $\mathcal{O}$. The circumcircle of each triangle does not contain any point in $\mathcal{O}$, and the partition tends to avoid sliver triangles. More importantly, Delaunay triangulation has the ability to patch new points without re-partitioning the whole space \cite{guibas1992randomized} \cite{buffa1992obstacle}. 

\begin{figure}
 \centering
 \includegraphics[width=1\linewidth]{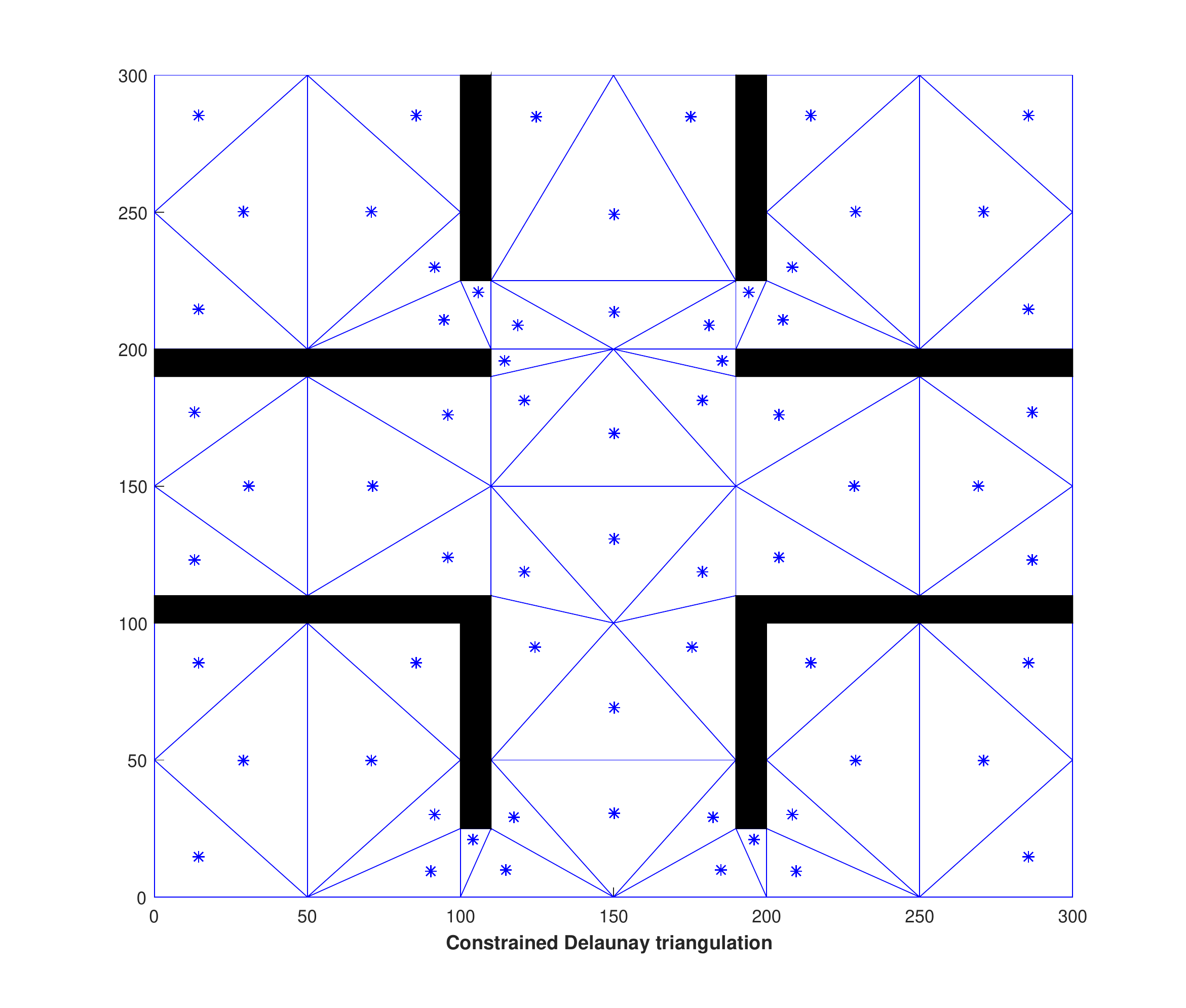}
 \caption{Constrained 2-D Delaunay triangulation.}
 \label{del}
 \vspace{-5.5mm}
\end{figure}

Without loss of generality, we assume that the obstacles region $\mathcal{X}_{obs}$ can be represented by a polygon, where the vertices of all obstacles, denoted by $v_{obs}\in V_{obs}$, are used in 2-D Delaunay triangulation. Using all vertices in $V_{obs}$ and several extra points in the environment to make sure the obtaining graph is not too sparse, the 2-D Delaunay triangulation partitions the working space with multiple triangles. We denote the center for each triangle $i\in [1,N]$ as $v_i$ where $N$ is the number of triangles after Delaunay triangulation. We define the edge $w_{i,j}=||v_i-v_j||$ if triangle $i$ and triangle $j$ share one edge. Then we construct a graph $\mathcal{G}=(\mathcal{V},\mathcal{W})$ for the environment where $\mathcal{V}$ is the set containing all $v_i$ and $\mathcal{W}$ includes all corresponding edges $w_{i,j}$. For the environment shown in Fig. \ref{comm}, considering obstacles (walls) and several extra points, we apply a constrained 2-D Delaunay triangulation and show results in Fig. \ref{del}. The constrained 2-D Delaunay triangulation \cite{peterson1998computing} is implemented by defining a boundary from the set of data points. In our case, the nodes in $V_{obs}$ are set to be the boundary. In order to include more details of the environment, we add more points for 2-D Delaunay triangulation. The blue stars in Fig. \ref{del} are the center of each triangle partition. 

\begin{remark}
In this paper, we assume the initial layout of the indoor environment is known {\it a priori}, which provides for building the initial 2-D Delaunay triangulation. It is worth pointing out the proposed high-level waypoints generation approach based on Delaunay triangulation is able to add new obstacles as robots exploring the environment. This is because Delaunay triangulation is capable of patching new points into existing partitions without regenerating the whole map. Once new obstacles are found, the waypoints generation layer will update the global map and provide a new set of waypoints for the local layer to follow.
\end{remark}

With the global map $\mathcal{G}$, the mission for the high layer is to generate a sequence of global waypoints $g_{i,q}, q\in[1,Q_i]\in \mathcal{V} $ to guide robot $\mathcal{R}_i$ to its final target, where $Q_i$ is the number of global waypoints for robot $\mathcal{R}_i$. A Dijkstra algorithm is employed to search the shortest path on the given graph, generating global waypoints for robots to follow.

\subsection{Synthesis of the Distributed MPC}
We wish to construct a distributed and online framework for communication-aware motion planning. To this end, we employ MPC as the basic framework so that the sub-problem for each robot is small enough to be solved online and the system is robust enough to deal with model uncertainty and external disturbance. However, the problem of completeness previously mentioned arises when MPC is introduced due to the limited planning horizon. We tackle this issue by adding a waypoint generation layer on the top. 

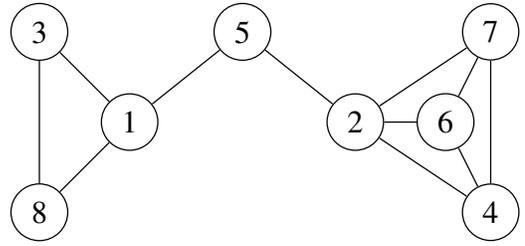
\begin{figure}
	\begin{center}
		\scalebox{1.2}{
			\begin{tikzpicture}
			\node[shape=circle,draw=black] (A) at (0,0) {8};
			\node[shape=circle,draw=black] (B) at (0,2) {3};
			\node[shape=circle,draw=black] (C) at (1,1) {1};
			\node[shape=circle,draw=black] (D) at (2.25,2) {5};
			\node[shape=circle,draw=black] (F) at (3.5,1) {2};
			\node[shape=circle,draw=black] (E) at (4.5,1) {6};
			\node[shape=circle,draw=black] (G) at (5,2) {7};
			\node[shape=circle,draw=black] (H) at (5,0) {4};
			
			\path [-] (A) edge node[left] {} (B);
			\path [-](B) edge node[left] {} (C);
			\path [-](A) edge node[left] {} (C);
			\path [-](D) edge node[left] {} (C);
			\path [-](D) edge node[above] {} (F);
			\path [-](E) edge node[above] {} (F);
			\path [-](E) edge node[above] {} (G);
			\path [-](G) edge node[above] {} (H);
			\path [-](E) edge node[above] {} (H);
			\path [-](F) edge node[above] {} (G);
			\path [-](F) edge node[above] {} (H);
			\end{tikzpicture}
		}
	\end{center}
	\caption{Neighbors of the robots.}
	\label{neighbor}
\end{figure}

Another strategy we employ for distributed and online computation is only considering the neighbors of a given robot in order to significantly reduce the size of each sub-problem. Since omitting distant robots is reasonable, we define the neighbor of each robot as those within a certain threshold distance of the given robot. We randomly assign each robot a unique priority in each planning period. A given robot can only plan in each cycle after all its neighbor robots with higher priority have finished planing. Fig. \ref{neighbor} illustrates one instance with randomly generated priorities. Robots that are directly connected are considered neighbors. For example, the robot with priority 2 plans first in its neighborhood in each planning cycle since it has the highest priority among all its neighboring robots. Then the robot with priority 4 will plan right after it.

We assume that global information, such as time, synchronization, and the states of neighboring robots, is available for each robot through communication base stations. The two-layer hierarchy planning algorithm is summarized in Algorithm \ref{algorithm}. For each waypoint in each planning period, robots formulate their own optimization problem in (\ref{J}) encoded as MILP, and run the MILP solver to find the control inputs within the planning horizon. After all the robots have planned, they implement the first step of the control inputs and move into the next period. The algorithm stops when all robots reach their goals or a time limit is reached.


\begin{algorithm}[ht]
	\SetAlgoLined
	\KwIn{Intial states and goal positions of each robot; 
	
	\qquad\quad Environment Structure;}
	
	\KwOut{Return states and control inputs of each robot}
	
	Initialization
	
	Generate graph $\mathcal{G}=(\mathcal{V},\mathcal{W})$ using constrained 2-D Delaunay triangulation based on the environment structure;
	
	\For{the Robot $\mathcal{R}_i$}{Generate a sequence of waypoints $g_{i,q}$ based on robots' initial and goal states and graph $\mathcal{G}$ using Dijkstra algorithm;
	
	Assign first waypoints $g_{i,1}$ to $\mathcal{R}_i$}

	\While{AgentSet $\neq \emptyset$ or $t \leq T_f$}{
		Randomly set a unique priority for each robot in AgentSet
		
		\For{$\mathcal{R}_i$, at time $t$}{
			Update the list of all robots' states  ${\bf x}_j^H$ and control inputs ${\bf u}_j^H$, $j\in\mathcal{P}$;
			
			\If{the Robot $\mathcal{R}_j$, $j\in\mathcal{P}$ is close enough to $\mathcal{R}_i$}{
				Add $\mathcal{R}_j$ into $\mathcal{R}_i$'s neighbour $\mathcal{N}_i$
			}
			
			Wait until all robots in $\mathcal{N}_i$ with higher priority than $i$ planned, then
			
			\For{$t\in[t,t+H]$}{
				Minimize the cost function $J_i$ in (\ref{J}) subjecting to corresponding constraints
				}
			
			Broadcast the results of ${\bf x}_i^H$ and  ${\bf u}_i^H$ to the whole group
			}
			
			Implement the first step of control inputs ${\bf u}_i^H$
			
			Global time $t=t+1$
			
		\If{A robot $\mathcal{R}_i$ $(i\in\mathcal{P})$ reaches its current assigned waypoints $g_{i,q}$}{Update waypoints to $g_{i,q+1}$}	
			
		\If{A robot $\mathcal{R}_i$ $(i\in\mathcal{P})$ reaches its final target}{
			Remove $\mathcal{R}_i$ from AgentSet
		}
	}
\caption{Distributed MPC Communication-aware Motion Planning with Completeness Guarantee}
\label{algorithm}
\end{algorithm}

\section{Simulation Results}
To validate our distributed co-optimization framework, we implemented our MPC based approach with high layer waypoints generation in MATLAB. The encoded MILP problem was modeled by AMPL, an algebraic modeling language for large-scale mathematical programming \cite{fourer1993ampl}, and solved by Gurobi, a commercial solver for MILP \cite{optimization2012gurobi}.

Fig. \ref{comm} is the basic setup for the simulation where four rooms are separated by walls with one open door for each room. We initially deploy six robots in the lower left room. The green areas in the other three rooms are the target areas that the robots want to explore. Each green area is required to be explored by two robots. Each room is equipped with one communication base station such that robots in the room can reach a base station. An extra base station is located at the center of the hall. Due to path loss of millimeter wave channel, it can only cover the blue area. In order to encode the given STREL formulas as MILP, we approximate the blue areas with n-sided polygons so that the service function $\lambda$ is a linear function. Three relay robots are deployed in the hall to help the other robots not in blue areas reach base stations.   

\begin{figure}
 \centering
 \includegraphics[width=1\linewidth]{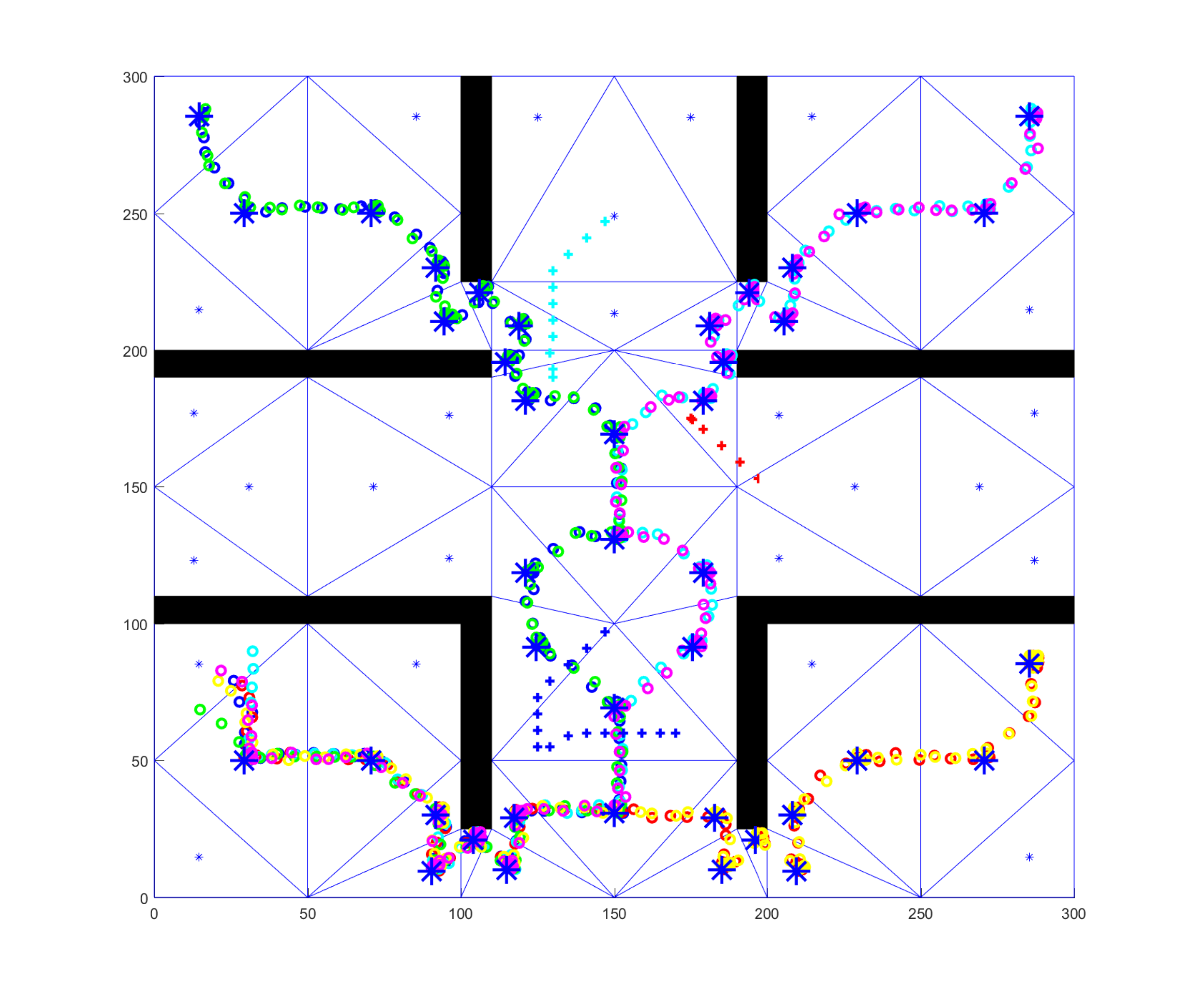}
 \caption{Communication-aware motion planning of networked mobile robots.}
 \label{sim1}
 \vspace{-5.5mm}
\end{figure}

\begin{figure}
 \centering
 \includegraphics[width=1.0\linewidth]{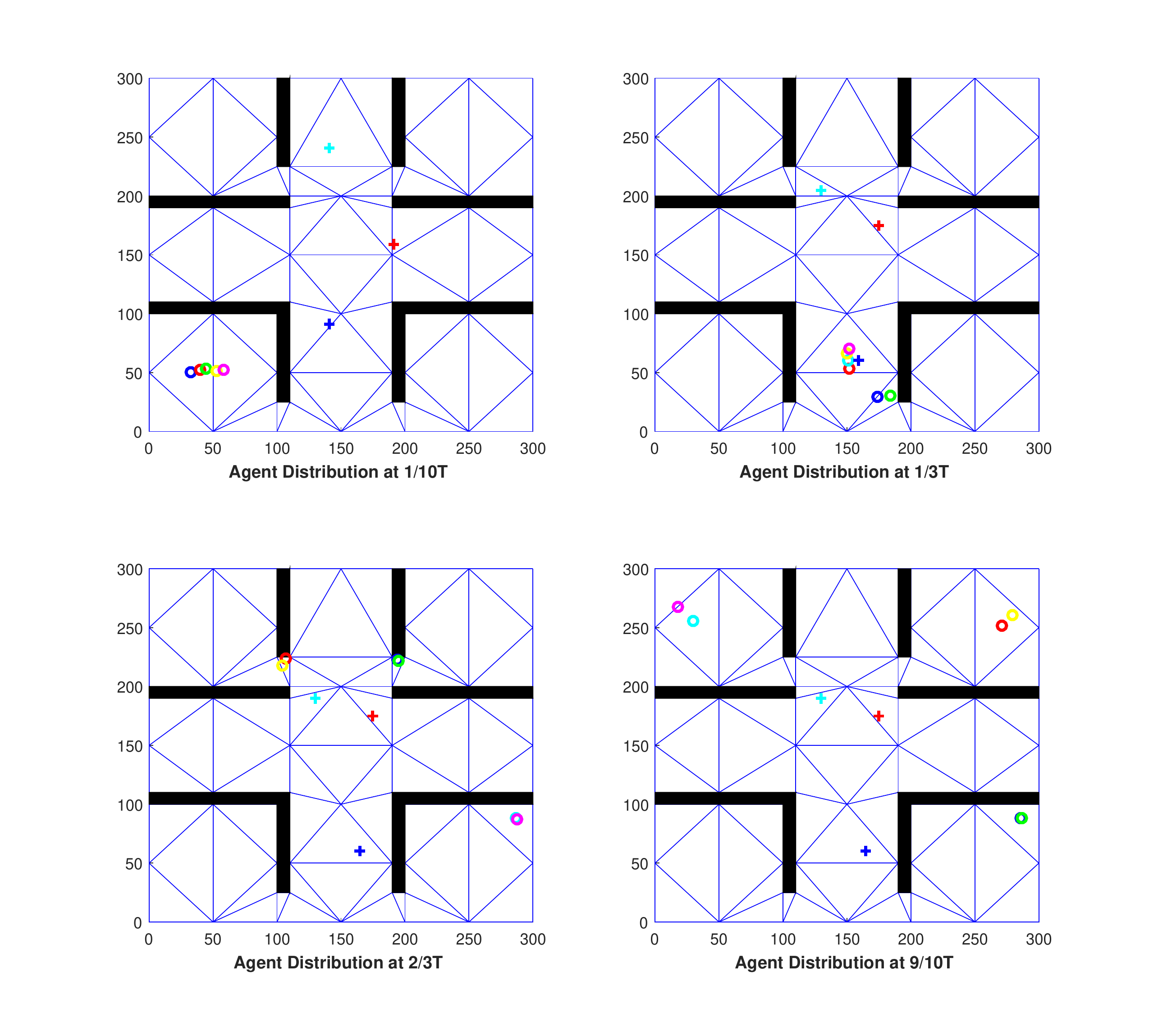}
 \caption{Distribution of the robots at different time steps.}
 \label{sim2}
 \vspace{-5.5mm}
\end{figure}

The dynamics of both exploring robots and relay robots, ruled by (\ref{agentdynamic}), are given by setting matrices $A_{i,d}$ and $B_{i,d}$ as
\begin{equation}
A_d=
\begin{bmatrix}
1 &0 &1 &0 &0 &0 \\
0 &1 &0 &1 &0 &0 \\
0 &0 &1 &0 &0 &0\\
0 &0 &0 &1 &0 &0\\
0 &0 &0 &0 &0 &0\\
0 &0 &0 &0 &0 &0
\end{bmatrix},
B_d=
\begin{bmatrix}
0.5 &0\\
0 &0.5\\
1 &0\\
0 &1\\
0 &0\\
0 &0
\end{bmatrix}.
\end{equation}
We choose $H=5$, $L=8$, $\Delta t=1$, $T_f=50$, $\eta =0.005$, $d_1=d_2=1$, and $\alpha=0.5$ with a $300m\times300m$ working space. Given vertices of the obstacles (walls) and several extra points, we apply the constrained 2-D Delaunay triangulation to partition the working space using multiple triangles. The partitions and their corresponding centers marked in small blue stars are shown in Fig. \ref{sim1}. The waypoint generation layer generates a sequence of waypoints marked by large blue stars in Fig. \ref{sim1} for robots to follow. Robots with missions are marked as circles, while relay robots are marked as crosses. Fig. \ref{sim1} demonstrates the whole process. Fig. \ref{sim2} shows the distribution of the robots at various time steps. From these results, we can see that robots are able to reach the target areas safely while satisfying the communication requirements specified by the STREL formula at the same time.  

The simulation was run on a PC with Intel core i7-4710MQ 2.50 GHz processor and 8GB RAM. The algorithm was run distributively among robots. Since the proposed distributed MPC framework significantly reduces computational complexity compared to the centralized co-optimization algorithm proposed in \cite{liuacc}, each robot can solve its own MILP problem in around 0.1 second at each planning period.

\section{Conclusion}
In this paper, we develop a distributed communication-aware motion planning framework for networked mobile robots via the application of MPC techniques. By utilizing STL and STREL formulas to specify the requirements for motion planning and communication, respectively, the formal specifications are encoded into MILP under distributed MPC. The proposed algorithm is able to find online control inputs for each robot distributively such that desired specifications and patterns can be satisfied, and hence demonstrates the ability of dealing with large-scale systems. STREL control synthesis is proposed by encoding STREL formulas as MILP, and control strategies can be obtained by solving the corresponding MILP. The communication and motion planning co-optimization framework is validated by the simulation of a networked mobile robots system.

\bibliographystyle{IEEEtran}
\bibliography{comm}
\end{document}